\documentclass[11pt]{book}
\usepackage{amsfonts}
\input epsf

\def\D{{\cal D}}
\def\cE{{\cal E}}
\def\I{{\cal I}}
\def\L{{\cal L}}
\def\M{{\cal M}}
\def\N{{\cal N}}
\def\O{{\cal O}}
\def\cR{{\cal R}}
\def\T{{\cal T}}

\def\Lieg{\mathfrak{g}}
\def\Lieh{\mathfrak{h}}
\def\Tr{\mathrm{Tr}}
\def\half{\frac{1}{2}}
\def\rank{\textstyle{\mathrm{rank}}}
\def\diag{\textstyle{\mathrm{diag}}}
\def\ad{\textstyle{\mathrm{ad}}}
\def\Aut{\textstyle{\mathrm{Aut}}}
\def\d{\partial}
\def\+{{+\!\!\!+}}
\def\pp{\mbox{\tiny${}_{\stackrel\+ =}$}}

\def\id{\hbox{1\hspace{-0.05in}1}} %blackboard bold 1
\def\bC{{{\mathbb C}\,}} %blackboard bold C
\def\bP{{\mathbb P}} %blackboard bold P
\def\bR{{\mathbb R}} %blackboard bold R
\def\bZ{{\mathbb Z}} %blackboard bold Z

\newcommand{\nc}{\newcommand}
\nc{\beq}{\begin{equation}}
\nc{\eeq}[1]{\label{#1}\end{equation}}
\nc{\ber}{\begin{eqnarray}}
\nc{\eer}[1]{\label{#1}\end{eqnarray}}

\begin{document}

\pagestyle{empty}

\frontmatter

\begin{titlepage}
                       
                                \hfill   hep-th/0305188\\

\begin{center}
  \hspace{0.5cm}

  \bigskip

  {\bf\Huge Superconformal D-branes \\
            and moduli spaces} \\
  \vspace{1cm}
  {\Large Cecilia Albertsson} \\
  \vspace{2cm}
    \begin{figure}[ht]
      \epsfxsize=9cm
      \centerline{\epsfbox{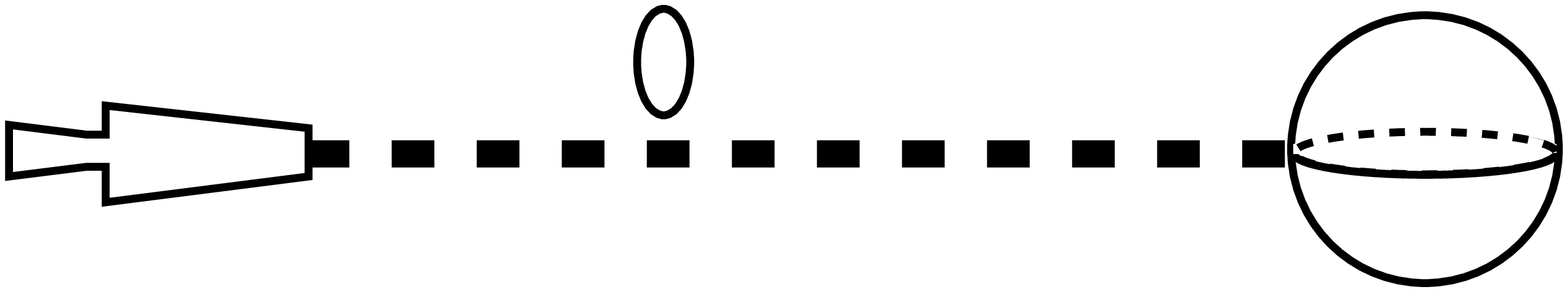}}
    \end{figure}
  \vspace{1.5cm}
  Doctoral Thesis in \\
  Theoretical Physics \\
    \begin{figure}[ht]
      \epsfxsize=2cm
      \centerline{\epsfbox{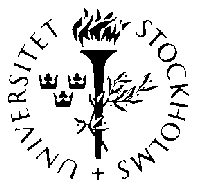}}
    \end{figure}
  Department of Physics \\
  Stockholm University \\
  2003
\end{center}
\end{titlepage}

\clearpage

\vspace*{\fill}

\noindent
Thesis for the degree of doctor of philosophy in theoretical physics\\
Department of Physics, Stockholm University\\
Sweden\\[\baselineskip]
\copyright \ Cecilia Albertsson 2003\\
ISBN 91-7265-630-1 (pp i-ix, 1-64)\\
Akademitryck AB, Edsbruk
  
\clearpage

\newpage

\section*{Abstract}

The on-going quest for a single theory that describes all the forces
of nature has led to the discovery of string theory. This is the only
known theory that successfully unifies gravity with the electroweak
and strong forces. It postulates that the fundamental building blocks
of nature are strings, and that all particles arise as different
excitations of strings. This theory is still poorly understood,
especially at strong coupling, but progress is being made all the
time. One breakthrough came with the discovery of extended objects
called D-branes, which have proved crucial in probing the
strong-coupling regime. They are instrumental in realising dualities
(equivalences) between different limits of string theory.
  
This thesis is concerned with dualities and D-branes.  First, it gives
a background and review of the first article, where we substantiated a
conjectured duality between two a priori unrelated gauge theories.
These gauge theories have different realisations as the worldvolume
theories on different D-brane configurations.  We showed that there is
an identity between the spaces of vacua (moduli spaces) arising in the
two theories, which suggests that the corresponding string theory
pictures are dual.
  
Second, we give a background and description of the analysis performed
in the remaining three articles, where we derived the most general,
local, superconformal boundary conditions of the two-dimensional
nonlinear sigma model. This model describes the dynamics of open
strings, and the boundary conditions dictate the geometry of D-branes.
In the last article we studied these boundary conditions for the
special case of WZW models.

\newpage

\section*{Acknowledgements}

First and foremost, I am immensely indebted to my advisor Ulf
Lindstr\"om, without whose help and guidance this thesis could never
have come into being.  He has always devoted ample time to answering
my questions and discussing problems whenever I needed to, often at a
mere moment's notice. I feel he always took great interest in my
progress, and offered lots of support and encouragement, especially at
times when I really needed it.

I have also greatly enjoyed working with both Bj\"orn Brinne and Maxim
Zabzine. It has been very stimulating, and their enthusiasm for
solving the mysteries of string theory is infectious to say the least.
In addition they contributed to a social and relaxed atmosphere during
my first two years in Stockholm, as did Tasneem Zehra Husain, who
also added colour and joy to our office in the new building.

Thanks also to Ron Reid-Edwards for providing the office entertainment
during my stay at Queen Mary, and to all the other people there
who made my visit an enjoyable one.

Ansar Fayyazuddin deserves special thanks for being the driving force
behind the study group, which I found incredibly useful, as does
Ingemar Bengtsson for encouraging me to apply to Fysikum in the first
place.

Last, but certainly not least, my eternal gratitude goes to mom and
dad for always believing in me, and for their unfailing support
through difficult times, well beyond the call of parental duty, by any
definition.

\newpage

\hspace{0.5cm}

\bigskip

\bigskip

\bigskip

\bigskip

\begin{flushleft}
\hspace{7cm} For Fay, \\
\hspace{0.5cm} \\
\hspace{7cm} who brightened my life \\
\hspace{0.5cm} \\
\hspace{7cm} for fifteen wonderful years \\
\end{flushleft}

\newpage

\addtocontents{toc}{\protect\thispagestyle{empty}}
\tableofcontents

\newpage

\vspace{0.5cm}
\noindent
{\bf \Large Accompanying papers}

\vspace{0.5cm}
\noindent
I: \emph{E$_8$ Quiver Gauge Theory and Mirror Symmetry}, \\
C.~Albertsson, B.~Brinne, U.~Lindstr\"om and R.~von Unge, \\
JHEP {\bf 05} (2001) 021, hep-th/0102038.
%%CITATION = HEP-TH 0102038;%%

\vspace{0.3cm}
\noindent
II: \emph{$N=1$ supersymmetric nonlinear sigma model with boundaries, I}, \\
C.~Albertsson, U.~Lindstr\"om and M.~Zabzine, \\
Commun.\ Math.\ Phys.\ {\bf 233} (2003) 3, 403-421, hep-th/0111161.
%%CITATION = HEP-TH 0111161;%%

\vspace{0.3cm}
\noindent
III: \emph{$N=1$ supersymmetric nonlinear sigma model with boundaries, II}, \\
C.~Albertsson, U.~Lindstr\"om and M.~Zabzine, \\
submitted to Commun.\ Math.\ Phys.\ (2002), hep-th/0202069.
%%CITATION = HEP-TH 0202069;%%

\vspace{0.3cm}
\noindent
IV: \emph{Superconformal boundary conditions for the WZW model}, \\
C.~Albertsson, U.~Lindstr\"om and M.~Zabzine, \\
JHEP {\bf 05} (2003) 050,
preprint no.\ UUITP-05-03, USITP-2003-02, \\
hep-th/0304013.
%%CITATION = HEP-TH 0304013;%%

\newpage

\mainmatter

\setcounter{page}{1}
\pagestyle{headings}

\chapter{Introduction}

\begin{flushright}
``This sort of thing has cropped up before, \\
and it has always been due to human error.'' \\
\hspace{0.5cm} \\
--- HAL 9000, \, in \, \emph{2001: A Space Odyssey} \\
\end{flushright}

\vspace{1cm}

Although the film's supercomputer was referring to a faulty
communications device, its statement turned out to carry universal
truth, much to the dismay of the ill-fated crew of \emph{Discovery}.
And not only was the fault indeed induced by humans, it was of a
fundamentally different nature than the crew initially thought.

This is characteristic of research in theoretical physics.  The quest
for an ultimate theory that will explain all the forces of nature in a
single, beautiful principle, is a typically human one.  Besides a
curiosity about the world around us, we are driven by our desire for
aesthetics and simplicity.  Sometimes so much so that we are tempted
to make oversimplified assumptions about nature, such as the
Aristotelian ``natural state,'' or Copernicus' circular planetary
orbits.  We may be unaware of the error, building entire theories on
our flawed axioms, and not until disaster strikes do we realise just
how fundamental a mistake we have made.

It is easy to make such mistakes because nature often turns out to be
much more peculiar than we ever imagined, displaying counterintuitive
effects like the particle-wave duality. At first, such weirdness
might be misconstrued as complications. But going along with it
usually in the end leads to an even simpler picture of the world,
bringing us closer to a unified theory of everything. So as a
physicist, one learns to accept ridiculous ideas just for the sake of
argument.

One such ``ridiculous idea'' constitutes the foundation of string
theory.

\section{Strings}

The idea is that all elementary particles are actually vibrating strings.

This was put forward by phycisists in the seventies, after failing to
use string theory to describe the strong interactions (quantum
chromodynamics does a better job of that).  It was realised that
string theory unifies gravity with the three forces described by
quantum field theory --- electromagnetism, the weak force and the
strong force.  Because the basic building blocks are one-dimensional
objects, strings, instead of zero-dimensional point particles, string
theory does not suffer from the divergences of quantum field theory.

The fact that strings are one-dimensional leads to a vast range of
possible string states.  Strings can be open (two ends) or closed
(ends joined), and they can oscillate in a multitude of ways.  The
different states that arise in this way correspond to different
particles.  That is, instead of the plethora of fundamental particles
in the Standard Model, we have only one type of fundamental object;
all matter can be explained as strings in different states.

But string theory is not a simple theory, in any sense of the word.
What keeps string theory a field in development is the fact that it is
technically very difficult, often impossible, to perform exact
calculations. One is frequently forced to approximate, or resort to
hand-waving. Furthermore, string theory is actually not a single
theory; it is \emph{five} different theories, all individually
consistent, but with different characteristics.  Some describe only
closed strings, others both open and closed, the strings may be
oriented or unoriented, and the theories have different symmetries,
etc.  At first sight these theories seem to be very far from anything
resembling realistic physics.  For one thing, they are in general
\emph{supersymmetric}, i.e.\ they demand the existence of
superpartners of all observed particles, none of which have shown up
in experiments as yet. Another nuisance is their requirement of no
less than ten spacetime dimensions to live in.

Nevertheless, in the course of time there has been an increasing
amount of order brought to this mess, in the form of symmetries and
dualities. It turns out that the five string theories are related to
each other via various dualities, so that they are manifestly
equivalent in different limits. One nice thing about this is that
computations that are difficult in one picture may become easier in
the dual picture. More interestingly, there is mounting evidence that,
at the end of the day, all these theories are just different limits of
one and the same underlying theory, our holy grail. This is why the
understanding of dualities in string theory is of paramount
importance.

\section{D-branes}

In the nineties, it was discovered that in addition to strings, string
theory contains other types of extended dynamical objects, called
Dirichlet branes (\emph{D-branes}). The name comes from their function
as hypersurfaces on which open strings can end --- an endpoint stuck
to a D-brane obeys Dirichlet conditions. They can be of any dimension
as long as it fits inside the ten dimensions of string theory.
D-branes play a crucial role in relating the different string theories
to each other, primarily via their transformation properties under
duality.

One way of dealing with the six surplus spacetime dimensions (since we
experience only four), is to compactify them on tiny spaces, like
wrapping a piece of paper around a pencil. From a distance the paper
then looks one-dimensional, and we are rid of one dimension.  Besides
reducing the number of dimensions, this technique has the advantage
that the ``compact'' part of the string theory shows up as matter in
the noncompact dimensions. We can thus construct the four-dimensional
theory of our choice by compactifying on the appropriate manifold.

In this context D-branes are useful for visualising where in the
ten-dimensional string theory our four-dimensional world fits. If we
choose a D-brane with four spacetime dimensions (a D3-brane), then the
part of string theory that lives on its \emph{worldvolume} would
describe the physics of the universe as we know it. Of course we would
need to break a lot of symmetries first, especially supersymmetry, but
in essence this is the picture.

\section{Outline}

This thesis is divided into two chapters; the first one
is concerned with Paper~I, and the second with Papers~II--IV.

\subsubsection{Moduli spaces}

The realisation of physics as the worldvolume theory of a D-brane is
the topic of Chapter~\ref{moduli}.  After providing some basics
concerning Lie algebras, we discuss the rather multifaceted background
of Paper~I. The centre of attention is the space of vacua
(\emph{moduli}) in the worldvolume theory of D3-branes, which splits
into different branches depending on the particular configuration we
are looking at. We explain how these branches of vacua arise, first
from a purely field-theoretical point of view, and then from a string
theory perspective. The object of Paper~I was to show the equivalence
between the moduli spaces of two different theories, the \emph{E$_8$
  quiver gauge theory}, and the \emph{E$_8$ Seiberg-Witten theory}, in
order to substantiate a conjectured duality between them. We define
these two theories and give a brief account of the method used to
compare the moduli spaces.

\subsubsection{Boundary conditions}

The dynamics of open superstrings is described by the supersymmetric
\emph{non-linear sigma model}, which is a field theory in two
dimensions. The domain of this model is the two-dimensional worldsheet
of the string, i.e.\ the surface that the string sweeps out as it
moves through spacetime. Since the string has two ends, this domain
has two boundaries, which by definition are attached to D-branes. So
studying boundary conditions of the sigma model is equivalent to
studying the geometrical properties of D-branes.

In particular, these conditions should be consistent with the way
D-branes transform under duality. For instance, \emph{T-duality}
changes the dimension of the D-brane, and the duality transformation
acting on the boundary conditions should yield the same result.  This
was our prime motivation in deriving the most general boundary
conditions possible, in Papers~II--III. More precisely, we derived
\emph{superconformal} boundary conditions, i.e.\ conditions for the
boundary to respect the super- and conformal symmetries that are
preserved in the bulk of the worldsheet.  Chapter~\ref{boundary}
provides some background to that derivation and explains how it was
done.  In Paper~IV we applied our analysis to the \emph{WZW model}, a
special case of the nonlinear sigma model; the last section is devoted
to that.

\section{About the thesis}

This thesis touches on many different topics, from F-theory to almost
product structures, all of them incredibly rich fields. We will not go
into great depth in any of these subjects, only give a brief review of
those aspects that are directly relevant to the accompanying papers. I
have tried to keep the discussion on a basic level, for the most part
assuming that the reader is not an expert. However, it has proven
unavoidable to sometimes state facts without justification, where an
explanation would be much too involved.  I have also tried to make
each chapter selfcontained, but inevitably, Chapter~\ref{boundary}
does make use of some concepts introduced in Chapter~\ref{moduli}.

Moreover, as frequently becomes apparent in string theory,
computational techniques, and even notation, are not of secondary
importance. This is definitely true in the work presented here, where
reaching the goal depended crucially on the method of getting there.
Despite this, we will not linger on technical details, only mention
briefly the approach taken in each case.

The interesting part is after all the conclusions.  They clarify but a
small fraction of the huge scientific effort which is string theory,
but I still believe that this theory is the path to follow in our
search for the ultimate principle.

Indeed, there is every indication that string theory is not merely a
misconception, due to human error.

\chapter{Moduli spaces}
\label{moduli}

\section{Introduction}

In this chapter we will be dealing with four-dimensional
supersymmetric gauge theories and their moduli spaces, realised as
worldvolume theories on D3-branes.  The exact worldvolume action
(i.e.\ including massive fields) on a D-brane is not known, although
considerable effort is being invested in finding it
\cite{Bilal2,Koerber}. So reliable analysis is possible strictly in
the low-energy effective theories, where only massless fields are
considered. Such theories are described by \emph{super-Yang-Mills}
theories, which will be our main concern here.

In particular, we are interested in the spaces of vacua in the
Yang-Mills theories. These are obtained by applying the \emph{Higgs
  mechanism} to the scalar potential; this mechanism renders gauge
bosons massive via spontaneous symmetry breaking, and is a candidate
for explaining the origin of mass. In $\N$=2 super-Yang-Mills theory
it gives rise to two moduli space branches, the \emph{Coulomb branch} and the
\emph{Higgs branch}. In string theory, this moduli space corresponds to
the space transverse to D-branes sitting in ten dimensions.

To construct four-dimensional theories from string theory, one usually
compactifies the ``superfluous'' dimensions on very small spaces so
that they become invisible in everyday, low-energy physics. These
compact spaces are subject to a set of restrictions in order that the
resulting theory be consistent; such spaces are known as
\emph{Calabi-Yau manifolds} \cite{Greene}.  A complex-one-dimensional
Calabi-Yau manifold is topologically always a torus, while in two
complex dimensions all Calabi-Yau manifolds are topologically
equivalent to the \emph{K3 surface} \cite{Aspinwall1}. The latter
space has orbifold singularities and is the one relevant
to us here. The great thing about string theory in this
context is that it remains well-behaved even when compactified on
singular spaces such as these orbifolds.

\subsection{Mirror symmetry}

It is believed that most, if not all, Calabi-Yau manifolds have an
associated \emph{mirror manifold}, i.e.\ a manifold whose complex
structure moduli are exchanged with the K\"ahler structure moduli as
compared to the original Calabi-Yau \cite{Aspinwall1}.  This is a very
important result since it has an implication that string theories
compactified on two mirror manifolds are equivalent (dual) to each
other.  As a consequence, if the moduli spaces of two conformal field
theories make up a mirror pair, then the corresponding theories are
dual to each other \cite{Greene,PolII}.

Intriligator and Seiberg \cite{IS} showed how a duality between two
different gauge theories in three dimensions corresponds to a mirror
symmetry between their respective moduli spaces. The two kinds of
theories are, on the one hand, three-dimensional ADE \emph{quiver
  theories} (as constructed by Kronheimer \cite{Kronheimer}), and on
the other hand $SU(2)$ gauge theory with ADE global symmetry
(\emph{Seiberg-Witten theory}). This mirror symmetry exchanges the
Coulomb branch of one theory with the Higgs branch of the other, and
vice versa. The two branches are in fact geometrically identical, but
the mirror exchange is interesting from a physics point of view. In
particular, the mass parameters (which are associated with complex
structure) of one theory are interchanged with the Fayet-Iliopoulos
parameters (associated with K\"ahler structure) of the other theory.

Since the four-dimensional versions of these theories are related to
the three-dimensional ones via compactification, one might suspect
that there is a similar mirror symmetry acting in four dimensions.  In
fact, a Higgs-Coulomb identity analogous to that of the
three-dimensional case was confirmed in \cite{E6}, for the $A_1$,
$A_2$, $D_4$ and $E_6$ four-dimensional theories.  The remaining two
of the strongly coupled superconformal $\N$=2 theories,
$E_7$ and $E_8$, were shown to also satisfy
such Higgs-Coulomb identities, in \cite{E7} and Paper~I, respectively.
Just like in three dimensions, the four-dimensional mirror symmetry
would provide a map between mass parameters of one theory and
Fayet-Iliopoulos parameters of the other.

One very useful consequence of such a mirror symmetry is that, since
the Coulomb branch receives quantum corrections but the Higgs branch
does not (due to $\N$=2 supersymmetry), quantum effects in one
theory arise classically in the dual theory, and vice versa. This
facilitates the analysis of nonperturbative phenomena enormously.
Although the $E_8$ theory does have some interest in itself, for
instance in explaining the $E_8$ gauge symmetry of the heterotic
string, the main motivation for establishing the moduli space
equivalence for $E_8$ was to complete the analysis for the whole
series of strongly coupled superconformal $\N$=2 theories.  Knowing
that there is a true duality between the quiver theory and the
Seiberg-Witten theory, one could use this to analyse the quantum
behaviour of physically more interesting theories such as $D_4$.

\subsection{Outline}

Clearly, some knowledge of simple Lie algebras is required, so we
start by listing some fundamental facts about these in
Section~\ref{Liealg}. We then move on to discuss $\N$=2
supersymmetric Yang-Mills theories in Section~\ref{YangMills}, showing how
the moduli space of vacua arises as a result of the Higgs mechanism.
Next, we define the two different gauge theories involved in the
duality discussed above, namely Seiberg-Witten theory
(Section~\ref{SWtheory}) and quiver theory (Section~\ref{Quiver}). The
latter section ends with a brief account of the computation done in
Paper~I.

Before proceeding, however, let us clarify a fundamental point, namely
the difference between ``perturbative'' and ``low-energy effective.''
A theory can be treated \emph{perturbatively} when the coupling
constant is so small that an expansion in powers of the coupling
constant is dominated by the first few terms.  On the other hand, a
theory is a \emph{low-energy effective} theory when any massive states
are so heavy compared to some fixed energy scale (e.g.\ the cutoff
scale in renormalisation) that they completely decouple from the
theory. Thus, for instance, the Seiberg-Witten theory is a strongly
coupled gauge theory where we have discarded all massive states and
retain only the massless ones.

With that, we are ready to embrace some representation theory.

\section{Lie algebras}
\label{Liealg}

Lie algebra theory is an essential instrument in a physicist's
mathematical toolbox. This is due to the close connection to vector
fields on manifolds, the most important example of which are those on
spacetime.  As the name suggests, Lie algebras are the algebras of \emph{Lie
  groups}, which by definition are groups endowed with the properties
of a smooth manifold. Examples of such groups are $GL(n,\bR)$, $SL(n,
\bR)$ and $SO(n, \bR)$.  The $C^\infty$ vector fields on such a
manifold form a Lie algebra.  We will be concerned only with
\emph{simple} Lie algebras, i.e.\ Lie algebras of finite dimension
greater than one and which contain no nontrivial ideals.  The complete
list of such algebras is not extensive, but the only ones relevant to
us are: $A_n$, $D_n$, $E_6$, $E_7$ and $E_8$. The first two correspond
to the groups $SL(n+1, \bC)$ and $SO(2n, \bC)$ respectively, whereas
the $E_n$ algebras correspond to three \emph{exceptional} groups,
defined e.g.\ in \cite{Fulton}. These groups are usually referred to
collectively as ADE symmetries and, somewhat confusingly, we sometimes
use the algebra notation to denote the corresponding groups.

In this section we give the standard definitions of some Lie algebra
objects that will be useful in the subsequent discussion.

\subsection{Definitions}

The Lie algebra $\Lieg$ is related to its Lie group $G$ by the
exponential map $\exp : \Lieg \rightarrow G$; i.e.\ $\exp (X) \in G$
for any element $X \in \Lieg$ .  This defines it as a representation
of its Lie group. That is, it is a homomorphism from $G$ to the group
of automorphisms of the tangent space of $G$ at the identity, $\rho: G
\rightarrow \Aut(T_e G)$ \cite{Fulton}. It comes equipped with a
\emph{Lie bracket}, defined as
\beq
[T^a,T^b] =f^{ab}_{\,\,\,\,\,\,\,c} T^c, 
\eeq{Liebracket}
where $T^a$ are the Lie algebra generators and $f^{ab}_{\,\,\,\,\,\,\,c}$
are structure constants. This bracket is a kind of product structure,
a bilinear form mapping any two elements in $\Lieg$ to a third.
As such it is used to construct the \emph{adjoint representation}
of the group $G$, defined as
$$
\ad_X Y \equiv [X,Y], \quad\quad X,Y \in \Lieg.
$$

The Lie bracket in principle defines the whole, rather rich, structure
of the algebra.  In particular, it defines the \emph{roots}, which are
eigenvalues of $\ad_h$, with $h$ any element in the \emph{Cartan
subalgebra} $\Lieh \subset \Lieg$. By a Cartan subalgebra we mean a
maximal abelian subalgebra such that the maps $\ad_h : \Lieg
\rightarrow \Lieg$ can be simultaneously diagonalised for all $h \in
\Lieh$. There are two types of roots, positive and negative,
which we denote by $w_+$ and $w_-$, respectively;
they are simply related by $w_- =  -w_+$.  Any root can be expressed as
a linear combination of a number of \emph{simple roots} with integer
coefficients. Positive roots are written with positive coefficients
and negative roots with negative coefficients in their
simple-root-decomposition.  The number of simple roots is equal to the
rank of the Lie algebra; for instance, $E_8$ has eight simple roots.

\subsection{Finding the roots}
\label{roots}

Roots satisfy a number of conditions which may be used to derive the
full set of positive roots \cite{Fulton}. As this was exploited in
Paper~I for the $E_8$ roots, we will demonstrate the procedure for
that particular case here,
but first we need to define another crucial ingredient in the Lie
algebra setup, the \emph{Cartan matrix}.  This is essentially a matrix
of inner products between simple roots.  More precisely, the
entries of the Cartan matrix $C$ associated with the Lie algebra $\Lieg$
are defined as
$$
C_{ij} \equiv 2 (\alpha_i, \alpha_j)/(\alpha_j,\alpha_j) ,
$$
where $\alpha_i$ are the simple roots, $i=1, ..., \rank(\Lieg)$,
and $(\, \cdot \, , \, \cdot \,)$ is the inner product on $\Lieg$.
The diagonal elements are always equal to 2, while the
off-diagonal elements are either zero or negative. For $\Lieg = E_8$
we have
\begin{eqnarray*}
C &=&  \left(
  \begin{array}{cccccccc}
     2 & -1 &  0 &  0 &  0 &  0 &  0 &  0 \\
    -1 &  2 & -1 &  0 &  0 &  0 &  0 &  0 \\
     0 & -1 &  2 & -1 &  0 &  0 &  0 &  0 \\
     0 &  0 & -1 &  2 & -1 &  0 &  0 &  0 \\
     0 &  0 &  0 & -1 &  2 & -1 &  0 & -1 \\
     0 &  0 &  0 &  0 & -1 &  2 & -1 &  0 \\
     0 &  0 &  0 &  0 &  0 & -1 &  2 &  0 \\
     0 &  0 &  0 &  0 & -1 &  0 &  0 &  2 \\
  \end{array}
  \right) .
\end{eqnarray*}
To find the full set of (positive) roots,
we first choose a root $w_0 = \sum_i m_i \alpha_i$ such that
$$
(w_0, \alpha_1) = 1, \quad\quad (w_0, \alpha_i) = 0, \,\, i \neq 1.
$$
Then the root properties imply that
$w_0 -\alpha_1$ is another positive root. Next,
we take the inner product of this new root with each of the simple
roots to see which one gives a positive result,
\begin{eqnarray*}
(w_0 - \alpha_1, \alpha_1) &=& 1-2 =-1, \\
(w_0 - \alpha_1, \alpha_2) &=& 0+1 =1, \\
(w_0 - \alpha_1, \alpha_i) &=& 0+0 =0, \quad\quad i > 2.
\end{eqnarray*}
Thus the next positive root is obtained as $w_0 -\alpha_1 - \alpha_2$,
and so on. One can show that the resulting set of roots from this
algorithm is the full set \cite{Fulton}; for $E_8$ we find 120
positive roots.

The reader may notice that we are dealing with \emph{weights} rather
than roots in Paper~I. The reason the above procedure is still
applicable is as follows.  Corresponding to each representation ${\cal
R}$ of a Lie algebra $\Lieg$, there is a set of weights, the
eigenvalues of $\Lieh$ in this representation. If ${\cal R}$ is the
adjoint representation $\ad_h$, then the weights are precisely the
roots of the Lie algebra. And since the fundamental representation of
$E_8$ is the same as the adjoint one, finding the weights of the
fundamental representation is in fact equivalent to finding the roots
of $E_8$.

\subsection{Dynkin diagrams}

All the structure and properties of any simple Lie algebra can be
encoded in a simple graph called a \emph{Dynkin diagram}. Such a
diagram consists of a number of nodes linked by edges. Each node
corresponds to a simple root, and the number of edges between each
pair of nodes reflects the value of the inner product between those
two simple roots. There is a direct relation between the entries of
the Cartan matrix and the number of edges $N_{ij}$ between the simple
roots $\alpha_i$ and $\alpha_j$, $N_{ij} = C_{ij} C_{ji}$.

\begin{figure}[ht]
  \epsfxsize=8cm
  \centerline{\epsfbox{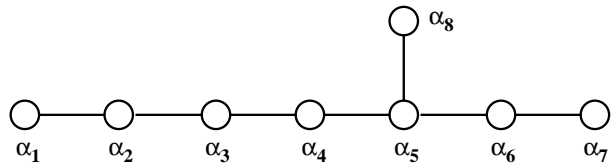}}
  \caption{\footnotesize The $E_8$ Dynkin diagram.
Each node corresponds to a simple root $\alpha_i$,
and the number of edges between each pair of nodes
is given by the Cartan matrix elements as
$N_{ij} = C_{ij} C_{ji}$\ ; compare to the $E_8$
Cartan matrix given in Section~\ref{roots}.}
  \label{E8dynkin}
\end{figure}

This kind of diagram is very powerful as an algebra representation
because, given a Dynkin diagram, you can recover the whole structure
of the corresponding Lie algebra. The $E_8$ Dynkin diagram is
shown in Fig.~\ref{E8dynkin}.

\section{$\mathbf{\N}$=2 super-Yang-Mills theory}
\label{YangMills}

Yang-Mills theory is essentially a synonym for nonabelian gauge
theory. That is, it is a field theory that is invariant under local
(or gauge) transformations, i.e.\ spacetime dependent transformations,
which form a nonabelian group.  In four dimensions such a theory can
be used to describe three of the four forces of nature at low
energy.\footnote{By ``low energy'' we mean the kind of energy scale
that is available to us in experiments. These energies can be as high
as several hundred GeV in present-day particle accelerators.} When the
theory has gauge symmetry group $SU(3)$, it describes the strong
interactions and is known as quantum chromodynamics (QCD). For
$SU(2)\times U(1)$ it unifies electromagnetic and weak interactions in
the electroweak theory. Together, QCD and the electroweak theory
constitute the Standard Model, which to date reproduces all known
experimental results of particle physics.

A gauge theory may possess other symmetries in addition to the gauge
symmetry. For example there may be a global symmetry group; that is,
the theory is invariant under some group of spacetime independent
transformations. The intuitive physical picture of such a global
symmetry is as a symmetry acting on added matter, for instance a
number of quarks in QCD.

Another symmetry example is supersymmetry, i.e.\ a symmetry between
bosons and fermions such that every boson is matched by a fermion (a
``superpartner'') with equal mass and charge. It may seem unmotivated
to introduce such a symmetry, since in experiments we have seen
neither spin-0 particles with the mass of an electron, nor massless
spin-$\half$ particles (``photinos''). But the idea is that the
low-energy world we live in has spontaneously broken supersymmetry,
while at sufficiently high energy we would see the supersymmetry
manifest. One promising clue that this might be the case comes from
the standard model coupling constants. The theoretical prediction is
that, without supersymmetry, they all are almost, but not quite,
equal, at around $10^{14}$ GeV, whereas inclusion of supersymmetry
makes them exactly identical, at an energy of around $10^{16}$ GeV.

If a nonabelian gauge theory is supersymmetric, it is called
super-Yang-Mills (SYM) theory. Such a theory is also invariant under
so-called \emph{R-symmetry}, which is essentially the symmetry group
transforming the different supersymmetry generators into each other.

We now define the precise type of SYM theories that interest us,
before embarking on an analysis of their vacuum states.

\subsection{Defining our SYM theory}

There are several parameters we need to specify in order to define
which particular type of SYM theory that we are interested in.  First,
SYM theory can be defined in any dimension up to ten (see
\cite{PolII}, Appendix~B), but the case relevant to us is the
four-dimensional one (as the low-energy effective action on a
D3-brane).

We also need to specify the number of supersymmetries, $\N$. Although
the worldvolume theory in Paper~I a priori has $\N$=$4$ (see
\cite{PolII}, Chapter~13), this supersymmetry is partially broken by
putting the branes on an orbifold singularity, and we end up with
$\N$=2. So we focus here on $\N$=2 SYM theory.

The next thing to specify is the field content.

\subsubsection{Field content}

A SYM theory contains a number of massless multiplets; the larger,
massive multiplets can always be decomposed into these.  The relevant
multiplet in ``pure'' $\N $=$ 2$ SYM theory (i.e.\ there are only
gauge field interactions in the theory, and no matter added by hand)
is a vector multiplet containing gauge fields $A^a_\mu$ ($a$ labels
the gauge group generators), two spinors $\lambda^a_\alpha$ and
$\chi^a_\alpha$, a complex scalar $\varphi^a$ and a real auxiliary
field $D^a$ (see \cite{PolII}, Appendix~B.2).

All these fields transform in the adjoint representation of the gauge
group.  This means that, under a transformation by an element $g$ of
the gauge group, a field $\phi$ transforms as $$
\phi \rightarrow  g \phi g^{-1}.
$$
On the other hand, it is said to transform in the \emph{fundamental}
representation if it transforms as
$$
\phi \rightarrow  g \phi, \quad\quad
\phi^{\dagger} \rightarrow  \phi^{\dagger} g^{-1} ,
$$
where $\phi^{\dagger}$ is the Hermitian conjugate of $\phi$.
Finally, $\phi$ is said to transform in the \emph{antifundamental}
representation of $G$ if it transforms as
$$
\phi \rightarrow \phi g^{-1}, \quad\quad
\phi^{\dagger} \rightarrow  g \phi^{\dagger} .
$$
Sometimes the representations are denoted by fat numbers, so that a
field transforming in the fundamental representation of, say, $U(3)$,
is said to transform as $\mathbf{3}$ (3 is the dimension of this
representation).  The antifundamental analogue is $\mathbf{\bar{3}}$.
Note that a field transforming in both $\mathbf{3}$ and
$\mathbf{\bar{3}}$ of $U(3)$ by definition transforms in the adjoint.
Moreover, if a field transforms under a product of groups, say $U(2)
\times U(3)$, as $(\mathbf{2}, \mathbf{\bar{3}})$, then we call it a
\emph{bifundamental} field. This notation will be relevant when we
discuss quiver gauge theories in Section~\ref{Quiver}.

We now add some fundamental matter to our pure $\N$=2 SYM theory.
More precisely, we introduce two \emph{hypermultiplets}, each of which
consists of a complex scalar field $\phi^i$ ($i=1,2$ labels the two
hypermultiplets), a spinor $\psi^i$ and a complex auxiliary
field\footnote{$F^i$ is called an auxiliary field because it has no
kinetic energy term. That is, its equations of motion are purely
algebraic and it can be expressed in terms of other dynamical fields.} 
$F^i$, and they all transform in the fundamental representation of the
gauge group.

\subsection{Spontaneous symmetry breaking}

Due to the shape of the potential in $\N$=2 SYM theory, the gauge
symmetry may be \emph{spontaneously broken}. This happens because,
instead of a unique vacuum with zero energy, there is a whole family
of vacua. Rather than being individually invariant under gauge
symmetry transformations, these vacua are transformed into each other.
The physical system will spontaneously choose one of the vacua, thus
breaking the gauge symmetry. This process goes by the name \emph{Higgs
  mechanism}, a physical example of which is superconductivity (see
e.g.\ \cite{Ryder}, Chapter 8).  The gauge-broken theory describes the
dynamics of the chosen vacuum field, which parameterises the
\emph{moduli space}, i.e.\ the space of vacua, of the theory.

To illustrate the principle of the Higgs mechanism, we now consider
the bosonic Yang-Mills theory with $SU(2)$ gauge symmetry.

\subsubsection{SU(2) bosonic Yang-Mills}

We first write down the Lagrangian and then define the constituent
fields.  The Yang-Mills Lagrangian, with one matter (complex scalar)
field $\phi$ transforming in the fundamental representation of the
gauge group, is (see \cite{Ryder}, Section 8.3)
\beq
\L_{YM}  = - (D^\mu \phi)^\dagger (D_\mu \phi) -
V(\phi^\dagger \phi) - \frac{1}{4} \sum^3_{a=1} F^a_{\mu\nu} F^{a\mu\nu} .
\eeq{LYM}
Since $\phi$ transforms in the fundamental
representation of $SU(2)$, we can write it as a
doublet,
$$
\phi (x^\mu) = \left( \begin{array}{c}
\phi^1 (x^\mu) \\
\phi^2 (x^\mu)
\end{array} \right) ,
$$
where $\phi^{1,2}$ are complex functions of the spacetime coordinates
$x^\mu$.

The field strength $F^a_{\mu\nu}$,
where $a$ labels the $SU(2)$ generators $T^a$ and $\mu, \nu$
are spacetime indices, is defined as
$$
F^a_{\mu\nu} \equiv \d_\mu A^a_\nu - \d_\nu A^a_\mu
-ig f^a_{\,\,\,\,bc} A^b_\mu A^c_\nu ,
$$
with $f^a_{\,\,\,\,bc}$ being the structure constants of the $SU(2)$
Lie algebra, cf.\ Eq.~(\ref{Liebracket}), and $g$ is a coupling constant.

We take the potential $V$ to be
\beq
V(\phi^\dagger \phi) \equiv \half \lambda^2
(\phi^\dagger \phi - \eta^2)^2 ,
\eeq{Vdef}
where the real number $\eta$ is the vacuum expectation value of
$\phi$.  In the bosonic theory there is nothing that forces us to
choose this particular potential, but in the presence of
supersymmetry there will be restrictions on $V$, and we choose
(\ref{Vdef}) to make the analogy as close as possible.

Finally, the covariant derivative of $\phi$ is defined as
$$
D_\mu \phi \equiv (\d_\mu - igT^aA^a_\mu)\phi
$$
(summation over $a$), where the generators $T^a$ are taken to be
in the fundamental representation of $SU(2)$; in terms of the Pauli matrices,
$T^a \equiv \sigma^a /2$.

\subsubsection{The Higgs mechanism}

We are interested in the classical vacua of this theory. These are
defined by the vanishing of the potential (\ref{Vdef}), so any vacuum
$\phi_0$ must satisfy $\phi^\dagger_0 \phi_0 = \eta^2$.  As mentioned
above, the $\phi_0$'s are not themselves gauge invariant, but $V$ is,
and there is a continuous family of vacua $\{\phi_0\} $ related by
$SU(2)$ transformations.  These vacua parameterise what is called a
``flat'' direction.

\begin{figure}[ht]
  \epsfxsize=6cm
  \centerline{\epsfbox{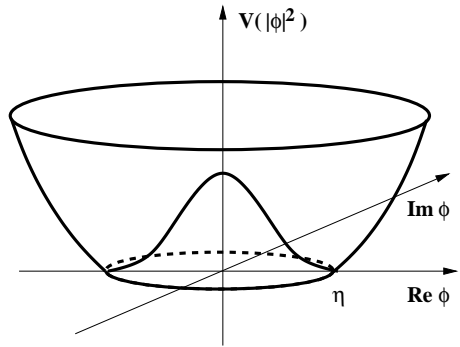}}
  \caption{\footnotesize The vacuum potential $V(\vert\phi\vert^2)$
of the Lagrangian (\ref{LYM}), as given by Eq.\ (\ref{Vdef}).
It has a continuous space of vacua at $\vert \phi
\vert^2 = \eta^2$, for which $V(\vert\phi\vert^2) =0$.}
  \label{mexican}
\end{figure}

To see where this terminology comes from, consider the $U(1)$ case,
where the vacua (which are now complex numbers) are related to each
other by a $U(1)$ transformation, $\phi_0' = e^{i\alpha} \phi_0$. As a
function of $\vert \phi \vert$, the potential $V(\vert \phi \vert^2)$
then has the shape of a mexican hat, with a minimum at $\vert \phi
\vert^2 = \eta^2$, where it vanishes, see Fig.~\ref{mexican}.  Moving
in a radial direction, either in towards the origin or away from the
origin, requires energy, like climbing up a hill.  But moving
along the minimum at a fixed distance from the origin requires no
energy since we are moving along the bottom of the ``moat'' --- this
is the flat direction, parameterised by the continuous spectrum of
vacua.

Returning to the $SU(2)$ case, we may now use the gauge freedom to
rotate $\phi_0$ to a basis where three of its four real components
vanish.  In other words, we are fixing the gauge by choosing a
particular vacuum.  The action is no longer invariant under $SU(2)$,
so we have broken the gauge symmetry; we have spontaneous symmetry
breaking.  However, note that the gauge symmetry is not completely
broken.  The Lagrangian (\ref{LYM}) is still invariant under $U(1)$
transformations, so the gauge group has been broken from $SU(2)$ down
to a $U(1)$ subgroup.

We are thus left with one real component in $\phi_0$, which we write as
a sum of a constant part $\eta$ and a nonconstant part $\gamma(x^\mu)$,
\beq
\phi_0 = \left( \begin{array}{c}
0 \\
\eta +  \gamma(x^\mu)
\end{array} \right) .
\eeq{phifixed}
If one inserts (\ref{phifixed}) into the Lagrangian (\ref{LYM})
and expands the latter, one finds that it contains mass terms for the
gauge bosons \cite{Maggiore}.
That is, there are quadratic terms of the form
$m^2_A A^a_\mu A^{a\mu}$, where the mass $m_A$ is proportional
to the parameter $\eta$.

In conclusion, spontaneous breaking of the gauge symmetry renders the
gauge bosons massive; in the $SU(2)\times U(1)$ case they combine into
the W- and Z-bosons (see \cite{Ryder}, Section~8.5).  The field
$\phi_0$ in this context is called a
\emph{Higgs field}, and its vacuum expectation value $\eta$ is the
\emph{Higgs mass}.

\subsection{The $\mathbf{\N}$=2 potential}

The Higgs mechanism generalises straightforwardly to the
supersymmetric theory. The difference is that supersymmetry imposes
constraints on the form of the action. For instance, the potential
cannot be arbitrary, and in addition the number of scalar fields is
restricted.

The action for $\N$=2 SYM theory is much more complicated than the
bosonic one, as it involves all the fields in the vector multiplet
($\varphi^a$, $\chi^a_\alpha$, $\lambda^a_\beta$, $A^a_\mu$, $D^a$)
plus any hypermultiplets $(\phi^i, \psi^i, F^i)$ added by hand, as
well as their interactions. In superspace formalism (see
Section~\ref{superspace} for details) this action looks simpler, but
we omit it here since we do not need to work with it explicitly. For a
pedagogical account of the $\N$=2 $SU(2)$ SYM action, see e.g.\ 
\cite{Bilal}.  Here it suffices to say that to find the potential for
the scalar matter fields (which we need for deriving the moduli
space), one expands the action in components and extracts all terms
containing the auxiliary fields $D^a$ and $F^i$. This yields a sum of
\emph{D-terms} and \emph{F-terms}, and then we integrate out $D^a$ and
$F^i$ by use of their equations of motion. The result is a potential
involving a sum of commutators between all scalar fields\footnote{Note
  that each scalar field is usually represented by a matrix, so the
  commutators become matrix commutators.} $\varphi^a$ and $\phi^i$.

\subsubsection{Fayet-Iliopoulos terms}

However, this is not the whole story.  In a supersymmetric gauge
theory, the precise form of the potential depends on the number of
$U(1)$ factors present in the gauge group. This is because
supersymmetry allows an extra term in the action when the gauge group
contains a $U(1)$ factor (see \cite{PolII}, Appendix~B). This extra term
is of the form $\xi^k D^k$, with $D^k$ being the $k$:th
$U(1)$-component of $D^a$, and $\xi^k$ is a real parameter, called the
\emph{Fayet-Iliopoulos term} (FI-term).  Thus the FI-parameter
corresponding to the generator $T^a$ is $\xi^a$ with $a$ running over
all the gauge group generators, but with $\xi^a \neq 0$ only for $a
= k$, where $k$ labels the $U(1)$ generators.

In $\N$=1 SYM there is only one, real FI-term for each $U(1)$
factor. $\N$=2 supersymmetry, on the other hand, allows \emph{three}
different FI-parameters for each $U(1)$; this is due to the $SU(2)$
R-symmetry that relates the two supersymmetries.  The three FI-terms
transform as a triplet under the $SU(2)$ R-symmetry \cite{Johnson},
and we denote them as a three-vector $\vec{\xi}^a$.

\subsubsection{The potential}

Let us finally have a look at the potential for the scalar fields
in our SYM theory:
\beq
 V =  \Tr ([\varphi, \varphi^{\dagger}]^2) +
\sum_{i=1,2} \phi^{\bar{i}} [\varphi, \varphi^{\dagger}] \phi^{i}
 + \sum^{\dim G}_{a=1}
\left( \Tr \left[T^a \cdot \vec{\mu} \right]  - \vec{\xi}^a \right)^2 ,
\eeq{SYMpot}
where a barred index implies Hermitian conjugate,
$\phi^{\bar{i}} \equiv \phi^{i\,\,\dagger}$.
The three-vector $\vec{\mu}$ is defined as
\beq
\vec{\mu} \equiv \Tr \left( \phi^\dagger \vec{\sigma} \phi \right),
\eeq{mudef}
where $\phi$ is the quaternion of the $\phi^i$'s,
\beq
\phi = \left( \begin{array}{cl}
\,\,\,\, \phi^{1 \,\, \dagger} & \phi^{2 \,\, \dagger} \\
-\phi^2 & \phi^1
\end{array} \right) ,
\eeq{phidef}
and $\vec{\sigma} = (\sigma^1,\sigma^2,\sigma^3)$ are the Pauli
matrices.

Note that each of the $\phi^i$'s is a matrix of rank equal to that of
$G$.  Similarly $\vec{\mu}$ is a three-vector of matrices of rank
equal to $\rank (G)$; the trace in (\ref{mudef}) is in the 2$\times$2
basis of the Pauli matrices, not over the matrices $\phi^i$.  On the
other hand, the trace in the last term in Eq.\ (\ref{SYMpot}) is in
the $\rank (G)$ basis, and its effect is to project $\vec{\mu}$
explicitly onto the basis vectors $T^a$.

We now use the potential (\ref{SYMpot}) to find the vacuum moduli
space.

\subsection{The SYM moduli space}

To find the vacua we set $V=0$ and solve for the scalars.  In a way
analogous to the bosonic analysis, we may fix the gauge and give
nonzero expectation values to the scalars. We thus end up with a
low-energy effective theory with a gauge symmetry that is a subgroup
of the original gauge group $G$, and massless scalars constituting the
moduli space.

Note that although nonzero expectation values of the scalars break the
gauge symmetry, they leave the supersymmetry unbroken. This is because
any vacuum state with zero energy (which is by definition true for the
Higgs fields) is supersymmetric as a direct consequence of the
supersymmetry algebra \cite{Wess}.  So in the case at hand, the
gauge-fixed theory also is $\N$=2 invariant, just like its
$G$-symmetric parent.

What is the geometry of the moduli space? There are a few different
possibilities, depending on the FI-terms. We first study the case
where all the FI-terms vanish; it is then clear from (\ref{SYMpot})
that the potential cannot vanish when both $\varphi^a$ and $\phi^i$
take generic values. So we have two possibilities: $\varphi^a \neq 0$
and $\phi^i=0$ on the one hand, and on the other hand $\varphi^a=0$
and $\phi^i \neq 0$. Thus the moduli space consists of two distinct
spaces, or \emph{branches}.

In the first case, when all hypermultiplets are zero, we find a family
of vacua $\{\varphi_0\}$ that transform into each other under $G$.
Spontaneous symmetry breaking renders all but one gauge boson massive
and we are left with one massless scalar $\varphi_0$.  The moduli
space is then complex-one-dimensional, parameterised by the gauge
invariant quantity $u \equiv \langle \Tr (\varphi_0)^2 \rangle$.  This
space, induced by the vector multiplet, is called the \emph{Coulomb
  branch}.  It is required by $\N$=2 supersymmetry to be a \emph{rigid
  special K\"ahler} manifold \cite{Aspinwall2}; in a four-dimensional
theory it is $\bP^1$.

Keeping $\varphi^a= 0$ on the other hand, and giving expectation
values to $\phi^i$ defines another branch of the moduli space, called
the \emph{Higgs branch}, and $\N$=2 supersymmetry requires it to be
a \emph{hyperk\"ahler} manifold \cite{HKLR}. This is a
real-$4k$-dimensional ($k$ an integer) manifold with $Sp(k)$
holonomy.\footnote{The \emph{holonomy} of a manifold is the subgroup of
  $O(n)$ under which a vector transforms as it is parallel-transported
  around a closed loop on an $n$-dimensional manifold.} It comes
equipped with three complex structures and three \emph{moment maps}.

Actually, the Higgs branch in this particular case has singularities;
it is an \emph{orbifold} with fixed points. Thus it is not a manifold
in the strict sense; however, it is the singular limit of a
hyperk\"ahler manifold, and as such it possesses all the structure of
a smooth hyperk\"ahler manifold.  An example is the \emph{K3
  orbifold}, i.e.\ the orbifold limit of a compact
complex-two-dimensional hyperk\"ahler manifold with $SU(2)$
holonomy.  Or
rather, we will be interested in orbifolds of the form $\bC^2/\Gamma$
(where $\Gamma$ is a discrete subgroup of $SU(2)$), which may be
viewed as a local description of a K3 orbifold near one of its
singularities.

\subsection{Singularities}
\label{sings}

The appearance of singularities on the moduli space is due to the way
we discard massive fields in the low-energy effective theory of the
Higgs fields.  One can show that, if the gauge-fixed scalars (the
Higgs fields) are
% UTSKRIVET TILL HIT!!
inserted into the SYM action, most of the gauge
fields acquire masses proportional to the vacuum expectation values of
the Higgs fields \cite{Bilal}, in analogy with the bosonic case.
These gauge fields may then be neglected in the low-energy effective
theory describing the gauge-fixed (massless) scalars; we say that the
gauge bosons have been ``higgsed away.''

The fact that the gauge-fixed theory includes only the massless fields
in the bulk (away from the origin of the moduli space) means that the
moduli space contains a singularity at the origin. The reason is that,
as the Higgs masses (the vacuum expectation values) approach zero, the
formerly massive fields become massless, and thus become relevant in
the theory. Therefore the low-energy effective bulk theory cannot be
accurate near the origin.

This is true classically for both branches; there is a singularity at
the origin of the Higgs branch which coincides with the singularity at
the origin of the Coulomb branch. However, when we pass to the quantum
level, the singularity of the Coulomb branch splits into several
separate singularities depending on the gauge group \cite{Lerche}.
The physical interpretation of the quantum singularities is not as
straightforward as in the classical case (gauge bosons becoming
massless), but for $SU(2)$ SYM it was shown in \cite{SW1} that two
singularities arise on the Coulomb branch, and that they correspond to
a pair of dyons (bound states of electric and magnetic charges)
becoming massless (see also \cite{Bilal}). Due to $\N$=2
supersymmetry, the Higgs branch receives no quantum corrections
\cite{SW1,Argyres1}.

\subsection{Nonzero FI-terms}

Continuing our investigation of the vacuum moduli space, it remains to
see what happens when all of the FI-terms $\vec{\xi}^k$ are nonzero
and generic. In this case the full potential (\ref{SYMpot}) can vanish
only if all the vector multiplets $\varphi^a$ are zero, which leaves
only the last term, involving the moment map.  The vanishing of this
term is commonly referred to as the \emph{D-flatness} condition. Since
$\varphi^a =0$, we again get a Higgs branch, except this time it looks
a little bit different.  It is again an orbifold, but with its
singularity resolved, or ``blown up.''  In geometric terms, this
blow-up is done essentially by replacing the singularity with a
connected union of intersecting two-spheres (two-cycles), and the
FI-terms parameterise the size of these spheres \cite{Aspinwall1}. The
resulting smooth space is then a true hyperk\"ahler manifold, called
an asymptotically locally Euclidean (ALE) space.

The FI-terms are in fact the \emph{periods} of the hyperk\"ahler
structures. This means that, if we represent the latter as a triplet
of two-forms comprising one K\"ahler form $J$ and two complex forms
$\omega$ and $\bar{\omega}$, then they are related to the triplet of
FI-parameters as \cite{Aspinwall1}
$$
\xi^k_1 = \int_{\Omega_k} J , \quad\quad
\xi^k_2 = \int_{\Omega_k} \omega , \quad\quad
\xi^k_3 = \int_{\Omega_k} \bar{\omega} .
$$
Here $\Omega_k$ is one of the two-cycles used to blow up the
singularity.  Because of this relation we see that the moment maps
$\vec{\mu}$ play a crucial role in resolving the Higgs branch
singularity, via the D-flatness condition \cite{HKLR}.

We summarise the moduli space in Table~\ref{t:branches}.
\begin{table}[ht]
  \begin{center}
    \begin{tabular}[]{c|c@{\hspace{1cm}}c}
      & Coulomb & Higgs \\ \hline
      $\vec{\xi}^k$ & \hspace{0.15cm} $0$ & \hspace{0.17cm} $0$ \hspace{0.2cm}
$\vert$ \hspace{0.1cm} $\neq 0$ \\
      $\varphi^a$ & $\neq 0$ & \hspace{0.16cm} $0$ \hspace{0.2cm}
$\vert$ \hspace{0.5cm} $0$ \\
      $\phi^i$ & \hspace{0.2cm} $0$ & $\neq 0$ \hspace{0.2cm}
$\vert$ \hspace{0.2cm} $\neq 0$ \\
Geometry & \hspace{0.2cm} $\bR^2$ & $\bC^2/\Gamma$ \hspace{0.06cm}
$\vert$ \hspace{0.2cm} ALE \\ \hline
    \end{tabular}
    \caption{\footnotesize
      The two vacuum moduli branches obtained by setting
the potential (\ref{SYMpot}) to zero.
For zero FI-terms the Higgs branch is an orbifold
$\bC^2/\Gamma$, whereas generic FI-terms resolve the singularity
to yield an ALE space.}
    \label{t:branches}
  \end{center}
\end{table}

Strictly speaking there is also the possibility of having only some of
the $\vec{\xi}^k$ vanish while the rest take generic values. Then a
zero potential does allow both $\varphi^a \neq 0$ and $\phi^i \neq 0$
simultaneously, for some combination of $a$'s and $i$'s
\cite{Johnson}. The resulting moduli space is then mixed, i.e.\ a
direct product of the Coulomb and Higgs branches. However, we will not
be concerned with this case here.

\section{Seiberg-Witten theory}
\label{SWtheory}

As explained in Section~\ref{sings}, the classical $\N$=2 Coulomb
branch has a singularity at the origin, and as we include quantum
corrections this singularity splits into several singularities.  Thus
the exact theory is fundamentally different from the classical
approximation.  This is an indication of the fact that perturbation
theory cannot be used in the region near the origin since the theory
is strongly coupled there.

However, it is in fact possible to determine the exact low-energy
effective theory, in the sense of calculating its exact complexified
coupling constant\footnote{The notation for the real and imaginary
  parts of $\tau$ is a convention.}  $\tau \equiv \frac{i}{g^2} +
\theta$ as a function of the Coulomb branch moduli. This was first
done by Seiberg and Witten \cite{SW1} for the $SU(2)$ symmetry-broken
$\N$=2 SYM theory (see \cite{Bilal} for a pedagogical review); we
therefore call this particular theory \emph{Seiberg-Witten theory}
(SW-theory).  The behaviour of $\tau$ at the singularities depends on
the type of each singularity, which in turn depends on the gauge group
and on any global symmetry of the unbroken SYM theory.

\subsection{Elliptic fibrations}

One may picture $\tau$ as defining the moduli of a two-torus, with the
real and imaginary parts giving the size of a one-cycle each.  Each
point on the Coulomb branch corresponds to a specific value of the
coupling constant, and therefore to a torus of specific dimensions.
This construction amounts to a \emph{fibration} of the torus over the
\emph{base space} constituted by the Coulomb branch. We call this
fibration the \emph{generalised Coulomb branch}, and it is a
complex-two-dimensional space described by the algebraic variety
\cite{Koblitz}
\beq
y^2 = x^3 + f(z) x + g(z).
\eeq{genCcurve}
Here $x$, $y$ and $z$ are complex variables and the functions $f$ and
$g$ are polynomials in $z$ whose degrees and coefficients depend on
the type of singularity we are dealing with. The variable $z$ is the
coordinate on the base space $\bP^1$ (the Coulomb branch), while $x$
and $y$ parameterise the tori.  This variety may be viewed as a
hypersurface embedded in $\bC^3$, in which case $x,y,z$ are the
complex coordinates on $\bC^3$.

Elliptic fibrations like these, where the fibres are tori
parameterised by a complex modulus $\tau$ and the base space is
$\bP^1$, exhibit singularities that have been classified according to
an ADE pattern \cite{Kodaira}.  There is a countably infinite number
of singularities at Im $\tau=\infty$; each such singularity is
of type either $A_n$ or $D_n$, for integer $n$.  In addition there are
seven singularities at finite values of $\tau$, of types
$A_0$, $A_1$, $A_2$, $D_4$, $E_6$, $E_7$ and $E_8$, respectively.

\begin{figure}[ht]
  \epsfxsize=4cm
  \centerline{\epsfbox{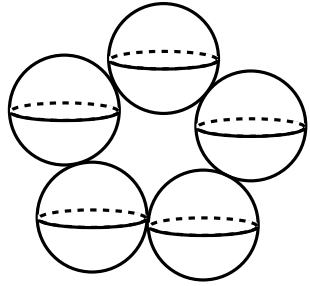}}
  \caption{\footnotesize The toroidal fibre at an $A_4$ singularity.
The torus is ``pinched'' in five places so that it becomes a necklace of
five two-spheres joined at points.}
  \label{necklace}
\end{figure}

When $z$ approaches one of the singularities on the Coulomb branch the
torus fibre degenerates in a specific way depending on the singularity
type. For instance, if the singularity is of type $A_n$, the torus is
``pinched'' in $n+1$ places so that it becomes a necklace of
two-spheres joined at points, as illustrated in Fig.~\ref{necklace}.
This singular hypersurface is then described by (\ref{genCcurve}) for
some specific polynomials $f$ and $g$; for $A_2$, for instance, $f$
goes to zero and $g= z^2$, so that the algebraic variety becomes $y^2
= x^3 + z^2$.

\subsection{Connection to physics}

This ADE classification of the fibration (\ref{genCcurve}) is a purely
mathematical result, but due to the interpretation of the torus
modulus $\tau$ as a coupling constant, it has inspired a line of
physics investigations that has proved very fruitful.  The idea is
that each of the singularities listed above corresponds to a
four-dimensional $\N$=2 SYM theory with \emph{global} symmetry
corresponding to the singularity type. In particular, the interesting
theories are the ones at strong coupling, i.e.\ at the seven
singularities at finite $\tau$ (Im $\tau = \infty$ corresponds to weak
coupling).

For instance, the original Seiberg-Witten theory, i.e.\ $\N$=2 SYM
with $SU(2)$ gauge group and four hypermultiplets (which has
$SO(8,\bC)$ global symmetry), fits nicely in this picture as the
strongly coupled theory at the $D_4$ singularity.  The $A_0$, $A_1$
and $A_2$ theories (i.e.\ they have global symmetries $SL(1,\bC)$,
$SL(2,\bC)$ and $SL(3,\bC)$) are obtained as certain limits of an
$SU(2)$ gauge theory with respectively one, two and three
hypermultiplets \cite{Argyres2,Eguchi}; these theories can be derived
from the $D_4$ theory.

The success of this correspondence thus far then prompted the
corresponding computation for the $E_6$, $E_7$ and $E_8$ theories,
i.e.\ theories with exceptional global symmetries\footnote{As
  Lagrangian descriptions do not exist for the exceptional theories,
  the authors of \cite{Minahan1,Minahan2,Noguchi} had to resort to
  more indirect methods of determining the generalised Coulomb
  branch.}  \cite{Minahan1,Minahan2,Noguchi}.  The existence of such
SW-theories\footnote{We extend the name SW-theory to
  include all the aforementioned strongly coupled ADE theories.} has
been shown also via compactifications of higher-dimensional gauge
theories \cite{Seiberg}.  The computation of exceptional varieties is
explained in detail in e.g.\ \cite{Lerche2}.

\subsection{Brane picture}

The picture of the coupling constant as a torus parameter has given
rise to the idea of \emph{F-theory} \cite{Vafa}. This is a conjectured
twelve-dimensional theory which, when compactified on a
four-dimensional K3 manifold, is equivalent to Type IIB theory
compactified on a two-dimensional manifold such as a sphere or a
two-torus.  The K3 manifold is a fibration of tori over the
two-dimensional manifold, and the coupling constant $\tau$
parameterises the fibre tori.

To see how this picture is relevant to us, we need to go into some
detail. Take the two-dimensional base space to be $\bP^1$,
parameterised by the complex coordinate $z$. Then the K3 manifold is
described by an algebraic variety of the form (\ref{genCcurve}) with
$f$ and $g$ being of degree eight and twelve respectively in $z$, and
it has 24 singularities.  These are determined as the zeroes of the
discriminant of the variety \cite{Koblitz}, and correspond in the IIB
picture to the positions on $\bP^1$ of 24 spacefilling 7-branes
(filling up the eight uncompactified dimensions) \cite{Johansen}.

When a specific combination of $(p,q)$ 7-branes (i.e.\ 7-branes on
which $(p,q)$-strings\footnote{A $(p,q)$-string is a bound state of
  $p$ fundamental strings and $q$ D1-branes.} can end) coincide at a
particular value of $\tau$, or equivalently, at a particular point on
$\bP^1$, then the gauge symmetry on the worldvolume of these 7-branes
is enhanced. Which particular gauge group we get depends on the number
and types of 7-branes involved \cite{Johansen,Gaberdiel}; the
different possibilities include the ADE groups and were categorised in
\cite{Noguchi}.

Next we do what Banks et al \cite{Banks} did and introduce a D3-brane
parallel to the 7-branes and located close to their position in
$\bP^1$. This ``probe technique'' is a popular approach to studying
the properties of string theory backgrounds; by ``probe'' we mean that
the D3-brane does not itself affect the background geometry. This is
admittedly an approximation, and backreaction has been taken into
account in e.g.\ \cite{Ansar2,Aharony}.

The probe provides an alternative picture of some of the physics on
the 7-branes, as features of the worldvolume theory living on the
D3-brane. This is a four-dimensional $\N$=2 SYM theory with a broken
gauge symmetry that becomes enhanced at the singularity.  Here it also
becomes superconformal, and acquires a global symmetry which is the
same as the gauge symmetry on the 7-branes\footnote{For $E_n$ global
  symmetry, this theory does not admit a Lagrangian description.}
\cite{Banks,Sen}.  Moreover, the hypermultiplet fields are strings
stretched between the D3-brane and the various 7-branes, whereas the
vector multiplets correspond to strings with both ends on the
D3-brane.

We have thus constructed a somewhat elaborate string theory setup in
order to obtain a brane interpretation of the globally symmetric,
superconformal theory. But it was worth the effort; we now have a
clear picture of the generalised Coulomb branch in terms of F-theory
--- it is just the K3 manifold on which the twelve-dimensional theory
is compactified (the ordinary Coulomb branch of the probe is the base
space $\bP^1$). In particular this has made it possible to find the
exact moduli space for strongly coupled SW-theories
\cite{Minahan1,Minahan2,Noguchi}.

Moreover, this setup is relevant here because now we have brane
pictures for both of the two gauge theories that Paper~I relates to.
The other theory, the \emph{quiver gauge theory}, which has a more
straightforward brane interpretation, is discussed in the next
section.  But first we remark that since, as was shown in \cite{E6,E7}
and Paper~I, there is a nontrivial identity between the moduli spaces
of these two different theories, we expect there to be some kind of
duality between the two string theory backgrounds.  However, this
turns out to be less than manifest, and attempts at finding such a
duality have failed thus far (see e.g.\ \cite{Feng}).

Let us now explain what we mean by a quiver theory.

\section{Quiver gauge theory}
\label{Quiver}

Quiver gauge theory is a well-established concept that frequently
crops up in string theory \cite{Johnson,Douglas}.  At first sight this
type of theory may seem anything but natural, as it involves a rather
specific gauge structure --- a product of unitary groups and matter
fields transforming according to a strict pattern as bifundamentals
under pairs of the constituent gauge groups. However, in the quest for
realistic physics based on string theory one must break both
supersymmetry and gauge symmetry in some way, and one of the most
straightforward procedures yields precisely what we call quiver
theory.

The idea is to start with Type IIB string theory in ten flat
dimensions (coordinates $x^0, x^1, ..., x^9$) and introduce a stack of
$N$ coinciding D3-branes.  We arrange these branes such that their
four-dimensional worldvolumes are aligned with the 0-1-2-3-directions
$x^0, ..., x^3$. The worldvolume supports a pure $\N$=$4$ SYM theory,
i.e.\ the only matter present is a vector multiplet containing three
complex (= six real) scalar fields transforming in the adjoint
representation of the gauge group. In the string theory picture these
scalars are the coordinates of the position of the branes along the
six dimensions transverse to the branes. As long as they all coincide,
the gauge symmetry is $U(N)$; this is due to the way in which the ends
of open strings are indexed (by \emph{Chan-Paton indices}) according
to which branes in the stack they are attached to (see \cite{PolI},
Section~6.5).

However, if the branes separate from each other, the gauge group is
broken down to some subgroup, since there will be fewer branes for
massless strings to end on. This is precisely the Higgs mechanism from
the point of view of the gauge theory; moving the branes corresponds
to giving expectation values to the scalar fields, which breaks the
gauge symmetry.

To break supersymmetry we make an orbifold out of the
6-7-8-9-dimensions by imposing an identification on the coordinates
under a group $\Gamma$. That is, two points are considered identical
if they differ only by a $\Gamma$-transformation.  Note that the
origin $x^\mu=0$ is fixed under $\Gamma$-transformations --- this is
the orbifold singularity, which will be blown up later.

Here we choose $\Gamma$ to be a discrete subgroup of $SU(2)$; the
motivation for this is that the resulting orbifold $\bC^2/\Gamma$
follows the same ADE classification as a K3 surface, and may be viewed
as a local description of a K3 singularity, at least near the origin.
As will become clear presently, this orbifold constitutes the Higgs
branch of the worldvolume theory, and as we mentioned in
Section~\ref{YangMills} it is required by $\N$=2 supersymmetry to be
just such an orbifold.

To summarise, we have the following configuration,
\begin{eqnarray*}
&& \begin{array}{|c|c|c|c|c|c|c||c|c|c|c|}
      \hline
\textstyle{\mathrm{Spacetime}} \,\, \textstyle{\mathrm{directions}}
  & 0 & 1 & 2 & 3 & 4 & 5 & 6 & 7 & 8 & 9 \\\hline
  |\Gamma| \,\,\, \textstyle{\mathrm{D3-branes}}
                  & \times & \times & \times & \times & \cdot
                  & \cdot & \cdot & \cdot & \cdot & \cdot \\\hline
  \textstyle{\mathrm{Orbifold}} \,\,\, \bC^2/\Gamma
                  &  &  &  &  & 
                  &  & \times & \times & \times & \times \\\hline
  \end{array}
\end{eqnarray*}
where crosses indicate the dimensions along which the brane or
orbifold extends, and a dot means the brane is pointlike in that
direction.

We now explore the consequences of the $\Gamma$-identification.

\subsection{Effects of orbifolding}

When we perform the orbifolding, the $U(N)$ gauge group breaks down to
a product of unitary groups, which we call $F$. One of the $\N$=$4$
complex scalars remains in the adjoint representation of the broken
gauge group. The other two scalars on the other hand transform as
bifundamentals according to a special pattern.  This rearrangement of
the matter fields breaks the supersymmetry down to $\N$=2; the
adjoint scalar becomes the scalar in the $\N$=2 vector multiplet,
while the two bifundamentals constitute the scalars of two
hypermultiplets.  Thus the worldvolume theory on the branes is now an
$\N$=2 SYM theory with two hypermultiplets.

It is easy to see explicitly how the orbifolding acts on the gauge
group and the scalars. A detailed account of the orbifolding procedure
is given in \cite{Johnson}, and we merely sketch it here. The
fields that we are interested in, namely the vector fields $A^a_\mu$
($a$ labels the gauge group generators), the complex 4-5-coordinate
$\varphi \equiv x^4 + ix^5$, and the complex 6-7-8-9-coordinates
$\phi^1 \equiv x^6 + ix^7$ and $\phi^2 \equiv x^8 + ix^9$, all arise
as massless excitations of open strings. We can therefore represent
them by matrices encoding their Chan-Paton indices. With $N$ branes
present, these matrices have dimension $N\times N$ with, in a suitable
basis, the ($i$,$j$) entry specifying whether or not the
string stretches between the $i$:th and $j$:th branes.

Here we take the number of branes to equal the order of the orbifold
group, $N= |\Gamma|$. Although the reason for this choice will become
clear later, we attempt to justify it already at this point.  Our aim
is to represent the action of $\Gamma$ on the open string sector, and
since there are $|\Gamma|$ distinct elements of the orbifold group, we
need $|\Gamma|$ different string states to represent them. Therefore
the strings need to be able to end on $|\Gamma|$ different branes;
these string states provide a faithful representation of $\Gamma$.

%\newpage

We use the following notation for the Chan-Paton matrices,
\ber
A^a_\mu & \rightarrow  & \lambda^a_V , \\
\varphi & \rightarrow & \lambda_I, \\
\phi^1 & \rightarrow &  \lambda^1_{II} , \\
\phi^2 & \rightarrow & \lambda^2_{II} .
\eer{ChanPaton}
In the unorbifolded theory, all these fields belong to the vector
multiplet and therefore transform in the adjoint representation of the
unbroken gauge group $U(|\Gamma|)$.  However, the requirement that
they be invariant under the orbifold group $\Gamma$ imposes
restrictions on the Chan-Paton matrices such that this is no longer
true.

If we denote by $\gamma_\Gamma$ the regular matrix
representation\footnote{The \emph{regular representation} of a group
  $G$ is the representation corresponding to the left action of $G$ on
  itself.  In matrix form it is a $|G|$$\times$$|G|$ block-diagonal
  matrix with each $k$-dimensional irreducible representation
  occurring $k$ times on the diagonal.} of $\Gamma$, then a field is
$\Gamma$-invariant if it commutes with $\gamma_\Gamma$. We therefore
impose
\ber
\lambda^a_V & = & \gamma_\Gamma \lambda^a_V \gamma^{-1}_\Gamma,
\label{lambdaV} \\
\lambda_I & = & \gamma_\Gamma \lambda_I \gamma^{-1}_\Gamma.
\eer{lambdaI}
For $\phi^{1,2}$, however, we need to take into account the fact that
they live along the orbifold directions.
This means that the invariance condition involves an extra
$\Gamma$-action on the doublet ($\phi^1$, $\phi^2$), via the 2$\times$2
matrix representation acting on the quaternion\footnote{For
explicit matrix representations of the ADE groups, see e.g.\
\cite{Johnson,Feng}.} (\ref{phidef}).
We call this matrix $G_\Gamma$, and find the following invariance
condition for $\phi^i$,
\beq
\lambda^i_{II}  = \left( G_\Gamma \right)^i_{\,\,j}
\gamma_\Gamma \lambda^j_{II} \gamma^{-1}_\Gamma .
\eeq{lambdaII}
It is now a matter of straightforward matrix algebra to derive the
form of the Chan-Paton matrices that satisfy Eqs.\
(\ref{lambdaV})--(\ref{lambdaII}), and the result is the
aforementioned factorisation
of the gauge group and the bifundamental structure. Some
explicit such calculations are shown in e.g.\ \cite{Feng}.

Note that the vacuum moduli space now consists of two branches, as
expected of a four-dimensional $\N$=2 theory.  The 4-5-space,
induced by the vector multiplet $\varphi$, constitutes the Coulomb
branch, and the orbifolded 6-7-8-9-space, induced by the
hypermultiplets $\phi^i$, is the Higgs branch.

So what about the Fayet-Iliopoulos terms? To find the corresponding
object in the string theory picture it is not sufficient to look only
at the open string sector. We need to include massless closed strings,
in particular the \emph{twisted} ones. That is, string states that are
invariant under the orbifold group only as long as they stay at the
singularity. These states enter the worldvolume theory of the
D3-branes in exactly the same way as FI-terms \cite{Johnson}.

\subsection{Quiver diagrams}

The resulting gauge structure has the interesting feature that it
falls into an ADE classification. That is, one may use the ADE
(extended) Dynkin diagrams to encode the transformation properties of
the bifundamentals. As explained above, these properties are directly
related to the group $\Gamma$. For instance, $\Gamma = \bZ_n$ breaks
the gauge group down to a direct product of $n$ $U(1)$'s and arranges
the hypermultiplets to transform as ($\mathbf{1}$, $\mathbf{\bar{1}}$)
under pairs of $U(1)$'s according to an $A_{n-1}$ pattern.

\begin{figure}[ht]
  \epsfxsize=8cm
  \centerline{\epsfbox{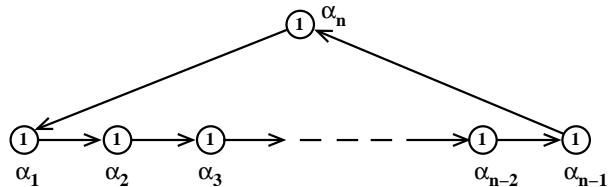}}
  \caption{\footnotesize The $A_{n-1}$ quiver diagram associated
with the $\bZ_n$ orbifold group.
The ``1'' in each node refers to the corresponding
 $U(1)$ factor in the quiver gauge group $U(1)^n$,
and the arrows indicate the transformation properties
of the bifundamentals. In this context the $\alpha_i$'s
denote FI-parameters.}
  \label{Anquiver}
\end{figure}

That is, we may represent the gauge structure by means of an $A_{n-1}$
extended Dynkin diagram, as shown in Fig.~\ref{Anquiver}. The nodes of
the diagram each corresponds to a $U(1)$ factor (i.e.\ a FI-term) and
the edges describe the bifundamentals with the arrow pointing towards
the antifundamental representation.  Similarly, for $\Gamma$ equal to
the dihedral group $\D_n$, we obtain a $D_{n-2}$ structure, and
$\Gamma = \T$, the tetrahedral group, yields an $E_6$ structure.  The
octahedral group, $\Gamma = \O$, gives $E_7$, and the icosahedral
$\Gamma = \I$ an $E_8$ structure. Thus there is a one-to-one
correspondence between the discrete groups $\Gamma$ and the ADE Lie
algebras.

That the quiver diagrams are \emph{extended} Dynkin diagrams means
that there is an extra $U(1)$ node compared to the ordinary Dynkin
diagram.  This extra $U(1)$ factor comes from the fact that there is a
trivial solution of (\ref{lambdaV}), namely $\lambda^0_V = \id$, the
identity matrix; the trivial solution corresponds to a $U(1)$ that
will always remain unbroken no matter what we do to the branes.  In
terms of the worldvolume gauge theory this means we can Higgs away all
the vector multiplets except the one corresponding to the trivial
$U(1)$. This $U(1)$ group is the gauge group in the worldvolume theory
of a single D-brane, so we see that the hypermultiplets now
parameterise the position in the orbifold of a single D3-brane.  From
the point of view of the covering space $\bC^2$, this brane is
actually a stack of $|\Gamma|$ \emph{fractional} branes, moving
simultaneously in such a way that they are images of each other under
$\Gamma$.

\section{The $\N$=2 Higgs branch}

The object of Paper~I was to show that the Higgs branch of the $E_8$
quiver theory is identical to the generalised Coulomb branch of $E_8$
Seiberg-Witten theory. There is no problem to do this in the singular
limit; then the quiver Higgs branch is just the orbifold
$\bC^2/\I$, which is described essentially by Eq.\ 
(\ref{genCcurve}) with $f(z)=0$ and $g(z)=z^5$.

The subtle difference is that the variables $x$, $y$ and $z$ are
$F$-invariants here, not $\I$-invariants, although they are
isomorphic to the latter \cite{Mumford}. To make the distinction
explicit, we call the $F$-invariants $X$, $Y$ and $Z$, and
find the variety \cite{E6}
$$
  Y^2+X^3+Z^5=0 
$$
for the $E_8$ quiver Higgs branch.

For nonsingular moduli spaces matters are more involved. We already
know the algebraic variety for the resolved SW-theory --- it is Eq.\ 
(\ref{genCcurve}), with \cite{Noguchi}
\ber
f(z) &=& w_2 z^3 + w_8 z^2 + w_{14} z + w_{20} \label{E8f} \\
g(z) &=& z^5 + w_{12} z^3 + w_{18} z^2 + w_{24} z + w_{30} .
\eer{E8g}
Here the coefficients $w_n$ are deformation parameters; when they are
nonzero, the singularity is deformed so that the space becomes smooth.

However, we did not know the explicit form of the resolved quiver
Higgs branch (the ALE space), so we computed it in Paper~I, in terms
of FI-parameters. The resulting variety was then brought, by means of
variable substitutions, to the form (\ref{genCcurve}) with $f$ and $g$
given by (\ref{E8f}) and (\ref{E8g}), except the coefficients in $f$
and $g$, which we called $\omega_n$, were now explicit polynomials in
FI-terms.  These polynomials may a priori be different from the
deformation parameters of the SW-theory, and the conclusion in Paper~I
was that they are in fact identical.

Before concluding this chapter, we briefly remark on the details
of the Higgs branch computation and comparison to the SW-variety.

\subsection{Comparing the moduli spaces}

To compute the Higgs branch we used so-called ``bug calculus,''
introduced in \cite{E6} and reviewed in \cite{ADEproc}.
It is essentially a technique to avoid
writing zillions of indices in computations that involve a lot of
fields. Polynomials in the ADE bifundamentals are represented by lines
drawn in quiver diagrams, and traces (invariants) correspond to closed
loops. These loops may then be manipulated subject to a set of
constraints that reflect the D-flatness conditions (i.e.\ the moment
map constraints), and we thus find the algebraic variety for the
quiver Higgs branch.

The comparison between the $E_8$ quiver Higgs branch and the $E_8$
SW-variety boils down to showing that our coefficients $\omega_n$ are
equal to the deformation parameters $w_n$ of Noguchi et al
\cite{Noguchi}.  The link between the two notations goes via the
simple roots. To see how, we introduce the \emph{characteristic
  polynomial},
$$
P^{\cR}_{G} \equiv \det (t \id - v \cdot H)
= \prod^{\dim \cR}_{k=1} (t - v_k),
$$
where $\cR$ is some representation of the group $G$, $t$ is a complex
parameter, and $v_k$ are the weights of the representation $\cR$.
The matrix
$v \cdot H \equiv \diag(v_k)$ is the matrix with the weights on
the diagonal and zeros otherwise, and $\id$ is the identity matrix. The
vanishing of $P^{\cR}_{G}$ encodes the same information about the
singularity in an elliptic fibration as the hypersurface
(\ref{genCcurve}) \cite{Lerche}.

In particular, it is convenient to use the characteristic polynomial
for computing the Casimir invariants of $G$, and this is what Noguchi
et al \cite{Noguchi}
did to express the $E_8$ Casimir invariants (= elements of $E_8$
that commute with all generators) in terms of the deformation
parameters $w_n$.  Their equations are easily inverted so as to
express the $w_n$'s in terms of Casimirs, which in turn are
polynomials in the weights $v_k$. And since the weights may be written
in terms of the simple roots as shown in Section~\ref{Liealg}, we obtain
the Casimirs as polynomials in the simple roots.  We thus have the
$w_n$'s expressed in terms of FI-parameters, hence they may be
explicitly compared with our coefficients $\omega_n$ (which were
defined as polynomials in FI-terms already from the beginning).

\chapter{Boundary conditions}
\label{boundary}

\section{Introduction}

This chapter is concerned with the dynamics of the ends of open
superstrings.  As an open string propagates through spacetime it
sweeps out a two-dimensional \emph{worldsheet}. The ends of the string
trace out one-dimensional paths, which constitute boundaries of the
worldsheet.  The motion of the string, and hence the shape of its
worldsheet, is dictated by equations of motion derived from a
two-dimensional field theory called the \emph{nonlinear sigma model}.
It is an action integral whose domain is the worldsheet, parameterised
by two coordinates: $\sigma$ along the string, and a time coordinate
$\tau$ along the direction of motion.  The \emph{target space} of this
integral is spacetime; that is, the dynamical fields in the action are
the vectors $X^\mu(\tau,\sigma)$, giving the position in spacetime of
the worldsheet point $(\tau, \sigma)$. Thus the dynamics of the string
is described by the equations of motion for $X^\mu$.

\begin{figure}[ht]
  \epsfxsize=6cm
  \centerline{\epsfbox{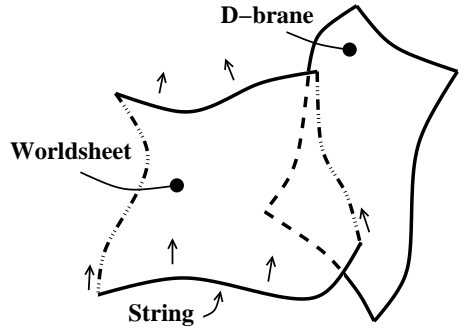}}
  \caption{\footnotesize An open string propagating in spacetime
sweeps out a two-dimensional worldsheet.
The hypersurface to which its end is confined defines a D-brane.}
  \label{worldsheet}
\end{figure}

In particular, the end of the string moves according to the equations
of motion on the domain boundary (boundary conditions), which restrict
it to move on some hypersurface in spacetime. Since open strings are
by definition attached to D-branes, this hypersurface is a D-brane.
Thus, in defining the hypersurface where the end is allowed to move,
the boundary equations of motion are telling us what the corresponding
D-brane looks like, see Fig.~\ref{worldsheet}.

How restrictive these equations of motion are depends on the amount of
symmetry preserved on the boundary. We focus here on the minimally
($\N$=1) supersymmetric conformal nonlinear sigma model,\footnote{In
  Papers~II and IV we also made a sketchy analysis of the $\N$=2
  model. In this case a rich structure arises due to an ambiguity in
  choice of sign in the boundary conditions. The $\N$=2 model was
  studied in more detail in \cite{LZ1,LZ2}.} assuming that the
worldsheet bulk superconformal symmetry is preserved also on the
boundary (the D-brane).  We showed in Papers~II--IV that the boundary
conditions allowed by this assumption are more general than those
commonly used elsewhere --- the latter conditions are just special
cases.  Nevertheless, we will see that our conditions do impose some
restrictions on the properties of D-branes.

In deriving the boundary equations of motion, the naive approach would
be to do so directly from the action, by use of the principle of least
action (i.e.\ a perturbation of the action should vanish). This is
hazardous, as the result does not necessarily preserve the desired
symmetries.  One could in principle amend this by modifying the action
by extra boundary terms to make it superconformal on the boundary.
However, there is no systematic way to find those boundary terms,
except guessing them.  In Papers~II and III we therefore went for the
safer method of analysing the currents that correspond to the relevant
symmetries, requiring that they be conserved on the boundary. This
defines boundary conditions for the currents, from which we could
derive conditions on the brane.

Having thus obtained the minimal requirements for boundary
superconformal invariance in a general background, it is natural to
consider a special case of background. In Paper~IV we chose the ever
popular \emph{WZW model}, which is a nonlinear sigma model defined on
a group manifold, with chiral isometry currents
\cite{Gawedzki,Stanciu}. Here our boundary conditions imply a
surprisingly general gluing map between the isometry currents, the
geometrical implications of which remain unclear at the time of
writing.

\subsection{Outline}

To introduce some fundamental concepts pertaining to symmetries and
conserved currents, we begin in Section~\ref{Bosonic} by discussing
the bosonic nonlinear sigma model.  Then we define the supersymmetric
sigma model in Section~\ref{superspace}, explaining about superspace
and superfields, and how to derive the superconformal currents. In
Section~\ref{ansatz} we make an ansatz for the worldsheet fields
which, after introducing some necessary notation in
Section~\ref{structures}, we plug into the conservation laws for the
currents and obtain in Section~\ref{BCs} the complete set of
conditions for a superconformal D-brane, which we interpret in
geometrical terms. Prompted by similarities to the structures of
\emph{almost product manifolds}, we look at globally defined boundary
conditions in Section~\ref{Global}, drawing some conclusions about the
global embedding of D-branes in spacetime.  Finally in
Section~\ref{WZW} we apply our analysis to the WZW model, leading to
some interesting statements about gluing maps of group currents.

\section{The bosonic model}
\label{Bosonic}

Consider an open string of tension $T$ propagating on a spacetime
manifold that supports a general background two-tensor $E_{\mu\nu} (X)
\equiv G_{\mu\nu} (X) + B_{\mu\nu} (X)$. Here $G_{\mu\nu}$ is the
spacetime metric and $B_{\mu\nu}$ an antisymmetric B-field, both of
which may depend on the spacetime coordinates $X^\mu$. Then the
nonlinear sigma model for the string is (see \cite{GSWI}, Section~3.4)
\ber
\nonumber S &=& -\frac{T}{2} \int d\tau d\sigma \, \left[ \,
\sqrt{-g} \, g^{ab} \d_a X^\mu \d_b X^\nu G_{\mu\nu} (X) \right. \\
&& \left.
+ \, \epsilon^{ab} \d_a X^\mu \d_b X^\nu B_{\mu\nu} (X) \, \right] ,
\eer{sigmabos}
where $g_{ab}$ is the metric on the worldsheet, $g \equiv \det
g_{ab}$, and $\epsilon^{ab}$ is the worldsheet antisymmetric
tensor with $\epsilon^{\tau\sigma}$=$-\epsilon^{\sigma\tau}$=1
and $\epsilon^{\tau\tau}=\epsilon^{\sigma\sigma}$=0.

One may derive equations of motion for the string by varying
(\ref{sigmabos}) with respect to $X^\mu$. Since the string is open,
the domain of the integral has boundaries, which contribute boundary
terms to the equations of motion.  The dynamics of the ends of the
string is then determined by the vanishing of these boundary terms,
implying some nontrivial boundary conditions for the open string.

The boundary conditions are of two types; either the string is moving
freely along the $X^\mu$-direction --- \emph{Neumann} conditions ---
or it is stuck in that direction --- \emph{Dirichlet} conditions.  The
hypersurface to which the string's endpoint is confined, i.e.\ the
D-brane, thus extends along Neumann directions and is pointlike in
Dirichlet directions (see \cite{PolI}, Chapter~8).

However, the boundary equations of motion obtained in this way do not
necessarily preserve all the symmetries that are preserved in the
bulk. If our prime concern is that they do (which it is), then we are
better off analysing \emph{conserved currents}.

\subsection{Conserved currents}

The model (\ref{sigmabos}) is invariant under three different
symmetries: spacetime Poincar\'e transformations, worldsheet conformal
transformations (rescaling of the two-dimensional metric $g_{ab}$),
and worldsheet reparameterisation. The last symmetry allows us to
switch to \emph{lightcone} coordinates on the worldsheet, $\xi^{\pm}
\equiv \tau \pm \sigma$, and we use this basis henceforth.

Each symmetry corresponds to a conserved current, obtained by varying
the action with respect to the appropriate field. The requirement that
the action be invariant under this perturbation (in the bulk)
translates into a conservation law\footnote{The conservation
  law holds only up to equations of motion; we say that it holds
  \emph{on-shell}.}  saying that the current is divergence-free, i.e.\ 
there are no sources.  For example in the case of conformal
invariance, the corresponding current is the stress energy-momentum tensor
and is derived by varying (\ref{sigmabos}) with
respect to $g_{ab}$. Its components in lightcone coordinates are
\beq
T_{\pm\pm} = \d_{\pp} X^\mu \d_{\pp} X^\nu G_{\mu\nu},
\eeq{Tbos}
where $\d_{\pp}$ are the derivatives with respect to $\xi^{\pm}$, and
we have rescaled the tension to one.  The $T_{++}$ component depends
only on $\xi^+$, and is called the ``left-moving'' current, whereas
$T_{--}$ depends only on $\xi^-$ and is referred to as
``right-moving.''  Here the bulk conservation law takes the form
$\d_{=} T_{++} = \d_{\+} T_{--} =0$.

To ensure that also the boundary is conformally invariant, we need to
impose current conservation on the boundary separately. In general we
find the boundary condition for a given current $J^a_{\,\,\,b}$ by
using its associated charge $Q_b \equiv \int d\sigma
J^\tau_{\,\,\,b}$. This charge is conserved in the sense that it is
time-independent, $\d_\tau Q_b =0$ (see \cite{Ryder}, Section~3.2).
Thus, since $J^a_{\,\,\,b}$ obeys the conservation law $\partial_a
J^a_{\,\,\,b} =0$, we may write charge conservation as
\beq
0=\d_\tau Q_b = \int d\sigma \,\,\, \d_\tau J^\tau_{\,\,\,b}
=- \int d\sigma \,\,\, \d_\sigma J^\sigma_{\,\,\,b}
= - [ \, J^\sigma_{\,\,\,b} \, ]^{\sigma=\pi}_{\sigma=0} .
\eeq{charge}
Hence we obtain the boundary condition $J^\sigma_{\,\,\,b} =0$, for
$b=\tau, \sigma$. Applied to the stress tensor, the result is
\beq
T_{++} - T_{--} =0 ,
\eeq{TbosBC}
i.e.\ the left- and right-moving components of the stress tensor must be
equal.  Via the relation (\ref{Tbos}) between the stress tensor and
the worldsheet fields, we thus find a boundary condition relating the
left-moving fields $\d_{\+} X^\mu$ to the right-moving fields $\d_{=}
X^\mu$.

In conclusion, we have derived the condition for conformal invariance
on the boundary in the bosonic model. But we are interested in the
supersymmetric theory, which looks a little bit different.

\section{The superspace action}
\label{superspace}

The bosonic model (\ref{sigmabos}) has only the three symmetries
listed in Section~\ref{Bosonic}. If we want more symmetry the action
needs to be modified. In particular, to make it supersymmetric we have
to add superpartner fields that together with the bosonic worldsheet
scalars $X^\mu$ make up multiplets (cf.\ Section~\ref{YangMills}). We
thus add two worldsheet spinors $\psi^\mu_{\pm}$ and an auxiliary
(i.e.\ nondynamical) field $F^\mu_{+-}$.

The fields constituting a multiplet can be conveniently collected in a
single \emph{superfield}. The idea is to promote the ordinary
worldsheet to a \emph{superspace} by supplementing the bosonic
worldsheet coordinates $\xi^{\pm}$ with anticommuting coordinates
$\theta^{\pm}$.  Superfields are terminating polynomials in
$\theta^{\pm}$, with the multiplet fields as coefficients. For
example, a multiplet ($X^\mu$, $\psi^\mu_{\pm}$, $F^\mu_{+-}$) would
correspond to the superfield \cite{Wess}
$$
\Phi^\mu \equiv X^\mu + i\theta^+ \psi^\mu_{+} 
+ i \theta^- \psi^\mu_{-} + \theta^+ \theta^- F^\mu_{+-}  .
$$
Similarly, the worldsheet derivatives $\d_{\pp}$ in (the lightcone
version of) (\ref{sigmabos}) are extended by additional
``superderivatives'' $D_{\pm}$, and the action becomes an integral
over superspace rather than over ordinary space. Superspace comes
equipped with integration rules that render this integral equivalent
to the ordinary one \cite{Wess}.

The whole point of using superfield notation is that the
supersymmetric theory can be analysed in a much more compact way than
if we were to use the explicit ``component form.''  Instead of writing
out the kinetic terms in the action for all the multiplet fields
individually, we can simply replace the bosonic fields in
(\ref{sigmabos}) with superfields. In lightcone coordinates we thus
obtain
\beq
S = \int d^2\xi d^2\theta
D_+ \Phi^\mu D_- \Phi^\nu {\cE}_{\mu\nu} (\Phi),
\eeq{sigmasusy}
where ${\cE}_{\mu\nu}$ is the superfield whose lowest component is
the background tensor
$E_{\mu\nu}$.  Analysis of the superspace action can then be performed
in a way completely analogous to the bosonic case, using superspace
quantities instead of bosonic ones.

Without boundaries, the action (\ref{sigmasusy}) is $\N$=(1,1)
(globally) supersymmetric, i.e.\ it is invariant under two independent
supersymmetry transformations that transform the bosonic and fermionic
fields into each other.  One is parameterised by a left-moving
supersymmetry parameter $\varepsilon^+$, and the other by a
right-moving one, $\varepsilon^-$. These two parameters are a priori
independent of each other, and each of them is associated with a
conserved supersymmetry current; we denote these currents by
$G_{\pm}$.

In the presence of a boundary the supersymmetry parameters
and currents are subject to boundary conditions that relate left- and
right-movers, as we saw in Section~\ref{Bosonic}. The
supersymmetry parameters become identified up to a sign,
$\varepsilon^+ = \eta \varepsilon^-$ ($\eta \equiv
\pm 1$), reducing the $\N$=(1,1) symmetry to $\N$=1. For the
currents the boundary condition (\ref{charge}) is
\beq
 G_{+}-\eta G_{-} =0 ,
\eeq{Gcond}
which is the condition for the boundary to preserve
worldsheet supersymmetry.

The condition (\ref{Gcond}) together with the stress tensor condition
(\ref{TbosBC}) define the superconformal boundary conditions for the
classical open superstring. It is these two conditions that served as
our starting point in Papers~II--IV.  To derive the corresponding
boundary conditions for the worldsheet fields we need to write the
currents $T_{\pm\pm}$ and $G_{\pm}$ in terms of $X^\mu$ and
$\psi^\mu_{\pm}$.  In the bosonic theory the expression for the stress
tensor was (\ref{Tbos}); here the relation is more complicated,
involving also $\psi^\mu_{\pm}$. We now briefly explain how
to compute the superconformal currents from a \emph{locally}
supersymmetric sigma model.

\subsection{Finding the currents}

The conformal and supersymmetry currents can be viewed as components
of a superfield which we call the ``supercurrent,'' and which we
denote by $T_A^{\,\,\,B}$ (the indices $A,B$ run over the superspace
indices $\+,=,+,-$).  This is a kind of stress tensor for a locally
supersymmetric version of the nonlinear sigma model (\ref{sigmasusy}).

What one does is to ``gauge'' the model by replacing the flat
superderivatives $D_\pm$ with covariant ones, $\nabla_\pm$, as well as
introducing a supervielbein\footnote{\emph{Vielbeins} are orthonormal
  tangent vectors that may be used to go to a locally flat tangent
  space at a point (see \cite{GSWII}, Chapter~12).}
$\mathfrak{E}_M^{\,\,\,A}$ on superspace (see \cite{GSWI},
Section~4.3.4).  The action then becomes
$$
S=  \int d^2\xi d^2\theta\,\, \mathfrak{E} \,\,
 \nabla_{+} \Phi^\mu \, \nabla_{-} \Phi^\nu \, {\cE}_{\mu\nu} (\Phi),
$$
where $\mathfrak{E}$ is the determinant of the supervielbein.

Next we vary $S$ with respect to the vielbein components and obtain
the supercurrent, of which only two components do not vanish on-shell,
namely $T^{\,\,\,-}_{\+}$ and $T^{\,\,\,+}_{=}$ \cite{Bastianelli}.
These are expressions in covariant superderivatives of superfields,
and we revert to global supersymmetry by replacing the covariant
derivatives with flat ones again. Finally, we can extract the
components of the supercurrent as follows,
$$
G_{+} = T_{\+}^{\,\,\,-} \vert ,\quad\quad
G_{-} = T_{=}^{\,\,\,+} \vert ,
$$
$$
T_{++} = -iD_{+} T_{\+}^{\,\,\,-} \vert  ,\quad\quad
T_{--} = - iD_{-} T_{=}^{\,\,\,+} \vert .
$$
The result is a set of explicit expressions for the currents in
terms of worldsheet fields, which we omit here; they
are given in Paper~III.

It is now in principle straightforward to convert the boundary
conditions (\ref{TbosBC}) and (\ref{Gcond}) for the currents into
boundary conditions for the fields $X^\mu$ and $\psi^\mu_{\pm}$.  We
know that the boundary enforces relations between left- and
right-movers, so we can make an ansatz for the way in which the left-
and right-moving worldsheet fields are related to each other, and then
derive restrictions on this ansatz from the current conditions.  Thus
our plan of attack is to make the most general ansatz possible for the
worldsheet fields, plug it into the current conditions, and reduce
these conditions to an independent set of boundary conditions that we
can interpret.

\section{The ansatz}
\label{ansatz}

It turns out that the most general, local ansatz we can make in our
classical conformal theory is very simple for the fermions,
due to the absence of dimensionful parameters. We therefore start
with this ansatz and then derive the corresponding
bosonic relation by a supersymmetry transformation. A little
dimensional analysis reveals that the fermionic ansatz takes the
form (the sign $\eta = \pm 1$
is included merely for convenience)
\beq
 \psi_-^\mu = \eta R^\mu_{\,\,\nu} (X) \,\psi_{+}^\nu ,
\eeq{fermbc}
where $R^\mu_{\,\,\nu}$ is a general (1,1)-tensor defined on the
boundary, which may depend on the worldsheet scalar $X^\mu$ at that
point, but not on $\psi_{\pm}^\mu$.  The bosonic superpartner of
(\ref{fermbc}) is more complicated,
\beq
\d_= X^\mu - R^\mu_{\,\,\nu}\d_\+ X^\nu + 2i(P^\sigma_{\,\,\rho}
\nabla_\sigma R^\mu_{\,\,\nu} + P^\mu_{\,\,\gamma} G^{\gamma\delta}
H_{\delta\sigma\rho} R^\sigma_{\,\,\nu})\psi_+^\rho \psi_+^\nu =0 ,
\eeq{bosbc}
where we have defined $2 P^\mu_{\,\,\nu} \equiv \delta^\mu_{\,\,\nu} +
R^\mu_{\,\,\nu}$ ($\delta^\mu_{\,\,\nu}$ is the Kronecker delta),
the antisymmetric three-tensor $H_{\mu\nu\rho}$ is
the field strength (or \emph{torsion}) of the background B-field, and
$\nabla_\sigma$ is the Levi-Civita connection.  Note that
(\ref{bosbc}) is more general than the usual boundary conditions
adopted in the literature (see e.g.\ \cite{Ooguri}), in that there is
an extra two-fermion term allowed by superconformal
invariance.\footnote{But see also \cite{Alvarez}, where an extra
  two-fermion term is included.}  We showed in Papers~II and III that
this extra term vanishes only for very special cases.

The ansatz (\ref{fermbc}) encodes in a covariant form the standard
Neumann and Dirichlet boundary conditions that define a D-brane.
In the presence of a B-field the Neumann condition has an a
priori very general form, while the Dirichlet condition is much
simpler. If we choose a basis where the worldvolume coordinates of the
D-brane are aligned with the spacetime coordinates --- called
\emph{adapted} coordinates --- then the Dirichlet condition takes the
familiar form
\beq
\d_\tau X^i =0,
\eeq{XDcond}
stating that the position of the endpoint along the $i$:th
Dirichlet direction does not change with worldsheet time $\tau$, i.e.\ 
it is frozen in that direction.  The corresponding situation holds for
the spinors,
\beq
\psi_-^i = -\eta \,\psi_{+}^i .
\eeq{psiDcond}

At first glance $R^\mu_{\,\,\nu}$ seems a completely general object,
but we can actually say something about it already at this stage, by
going to adapted coordinates at a point. In this basis it should imply
(\ref{psiDcond}), so the Dirichlet-Dirichlet part is $R^i_{\,\,j}
= - \delta^i_{\,\,j}$.  On the other hand, the Neumann condition is
still very general ($m,n$ label Neumann directions),
\beq
 \psi_-^m = \eta R^m_{\,\,n} (X) \,\psi_{+}^n ,
\eeq{psiNcond}
for some Neumann-Neumann tensor $R^m_{\,\,n}$.  It is clear
that this tensor depends on the B-field, because for $B_{\mu\nu}=0$ we
expect (\ref{psiNcond}) to reduce to $\psi_-^m = \eta \,\psi_{+}^m$.
In fact, one may think of $R^m_{\,\,n}$ as \emph{defining} the
B-field;\footnote{Note that $B_{\mu\nu}$ is actually the gauge invariant
  combination of a truly background B-field and the field strength
  of the $U(1)$ gauge field on the D-brane.} we will see
later precisely how.

To summarise, if we write $R^\mu_{\,\,\nu}$ as a 2$\times$2 block
matrix, with the upper left block being the Neumann-Neumann part, the
lower right the Dirichlet-Dirichlet part, and the off-diagonal blocks
the mixed parts, we have
\beq
R^\mu_{\,\,\nu}= \left(
\begin{array}{cc}
R^m_{\,\,n} &  0\\
0 & -\delta^i_{\,\,j}
\end{array}
\right) .
\eeq{Rmatrix}

But we want to work in a basis-independent notation, so we would like
to write for example the Dirichlet condition (\ref{XDcond}) in a
covariant form.\footnote{We focus on the Dirichlet condition because
  of its simple form.}  To do this, we need to introduce some
structures on our spacetime manifold.  These structures are natural
from a physical point of view, but there is a rich mathematical
machinery associated with them, which will be tremendously useful in
writing down and interpreting the final boundary conditions.

\section{Structures on D-branes}
\label{structures}

We begin by defining a projector $Q^\mu_{\,\,\nu}$ on the worldsheet
boundary, which projects vectors on the D-brane onto the space spanned
by the Dirichlet directions.  That is, given a vector $X^\mu$ at some
point on the brane, $Q^\mu_{\,\,\nu} X^\nu$ is by definition normal to
the brane at that point.  If $X^\mu$ is invariant under
$Q^\mu_{\,\,\nu}$, $Q^\mu_{\,\,\nu} X^\nu = X^\mu$, then it is a pure
Dirichlet vector.  In this sense, $Q^\mu_{\,\,\nu}$ assigns a vector
space to each point $u$ on the brane, which we denote by $Q_u$.
Clearly the dimension of this space equals the rank of
$Q^\mu_{\,\,\nu}$. We call $Q^\mu_{\,\,\nu}$ a \emph{Dirichlet
  projector}, and we can use it to write the Dirichlet condition
(\ref{XDcond}) on the desired covariant form,
$$
Q^\mu_{\,\,\nu} \d_{\tau} X^\nu =0 .
$$

Similarly, we may define a \emph{Neumann projector}
$\pi^\mu_{\,\,\nu}$ complementary to $Q^\mu_{\,\,\nu}$, satisfying
$$
\pi^\mu_{\,\,\nu} Q^\nu_{\,\,\rho} =0 ,
\quad\quad \pi^\mu_{\,\,\nu} + Q^\mu_{\,\,\nu} = \delta^\mu_{\,\,\nu} .
$$
It projects vectors onto the tangent space of the brane (i.e.\ the
Neumann directions) at any given point $u$ on the brane, and thus
assigns to $u$ a vector space $\pi_u$ of dimension $\rank (
\pi^\mu_{\,\,\nu} )$.

\begin{figure}[ht]
  \epsfxsize=6cm
  \centerline{\epsfbox{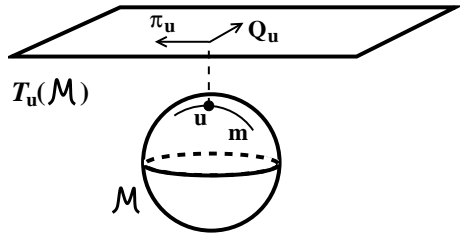}}
  \caption{\footnotesize The two projectors $\pi^\mu_{\,\,\nu}$
and $Q^\mu_{\,\,\nu}$ split the tangent plane $T_u(\M)$ of a manifold
$\M$ at the point $u$ into a direct sum of two orthogonal spaces
$\pi_u$ and $Q_u$. Under certain circumstances (see the text for details),
$\pi^\mu_{\,\,\nu}$ defines
the tangent space of a submanifold {\bf m} of $\M$.}
  \label{TM}
\end{figure}

The two spaces $Q_u$ and $\pi_u$ are subspaces of the tangent plane
$T_u(\M)$ of the spacetime manifold $\M$ at the point $u$, and they
are orthogonal to each other. That is, any vector in $Q_u$ is
orthogonal to all vectors in $\pi_u$ with respect to the metric
$G_{\mu\nu}$ on $\M$. In other words, $\pi^\mu_{\,\,\nu}$ and
$Q^\mu_{\,\,\nu}$ split $T_u (\M)$ into a direct sum
(see Fig.~\ref{TM}),
\beq
T_u (\M) = \pi_u \oplus Q_u .
\eeq{Tsplit}

It is worth emphasising that the only assumption we make from the
outset is that the target space is a smooth manifold equipped with a
(pseudo-) Riemannian metric and an antisymmetric two-tensor (the
B-field).  A priori we do not know anything about the properties of
the D-brane; it need not be a regular submanifold of spacetime, and
could be singular or ill-defined for all we know.  Our object is to
start from this unassuming standpoint and derive the properties
required of the brane by superconformal invariance.  It is thus by no
means certain at this point that $\pi^\mu_{\,\,\nu}$ defines a tangent
space corresponding to a submanifold of spacetime, although the vector
space $\pi_u$ is a subspace of $T_u (\M)$. However, we will see in
Section~\ref{Global} that, under certain circumstances, this follows
as a consequence of the superconformal boundary conditions.

We need one more object before writing down the boundary conditions.
Given two orthogonal
projectors $\pi^\mu_{\,\,\nu}$ and $Q^\mu_{\,\,\nu}$ we may define
another (1,1)-tensor,
$$
r^\mu_{\,\,\nu} \equiv \pi^\mu_{\,\,\nu} - Q^\mu_{\,\,\nu} ,
$$
which squares to the identity, $r^\mu_{\,\,\nu}
r^\nu_{\,\,\rho} = \delta^\mu_{\,\,\rho}$,
since $\pi^\mu_{\,\,\nu}$ and $Q^\mu_{\,\,\nu}$ are orthogonal and
each of them squares to itself.  Alternatively, $\pi^\mu_{\,\,\nu}$
and $Q^\mu_{\,\,\nu}$ can be written in terms of $r^\mu_{\,\,\nu}$,
$$
\pi^\mu_{\,\,\nu} = \half \left( \delta^\mu_{\,\,\nu}
 + r^\mu_{\,\,\nu} \right) ,
\quad\quad
Q^\mu_{\,\,\nu} = \half \left( \delta^\mu_{\,\,\nu}
- r^\mu_{\,\,\nu} \right) .
$$

The three structures $\pi^\mu_{\,\,\nu}$, $Q^\mu_{\,\,\nu}$ and
$r^\mu_{\,\,\nu}$, although very simple and intuitive in essence, were
crucial in deriving and understanding the superconformal boundary
conditions in Papers~II and III.  Note that initially they are defined
only at a point, the endpoint of the string. However, we will see that
superconformal invariance allows us to extend them to a neighbourhood,
thus providing some information about the local smoothness of the
brane.  In fact, we will even define them globally (i.e.\ at every
point in $\M$), just to see what happens. One can then draw some
conclusions about the way in which the D-brane is embedded, globally,
in spacetime, rather than merely looking at a small coordinate patch
of the brane. But before exploring such possibilities, let us finally
write down the boundary conditions explicitly.

\section{The boundary conditions}
\label{BCs}

Using the ansatz (\ref{fermbc}) and (\ref{bosbc}) in the current
conditions (\ref{TbosBC}) and (\ref{Gcond}), we found in Paper~III
the following three boundary conditions,
\ber
&& R^\mu_{\,\,\rho} G_{\mu\nu} R^\nu_{\,\,\sigma} =G_{\rho\sigma},
\label{RgR} \\
&& \pi^\rho_{\,\,\sigma} E_{\nu\rho} \pi^\nu_{\,\,\gamma}
= \pi^\rho_{\,\,\sigma} E_{\rho\nu} \pi^\nu_{\,\,\lambda}
R^\lambda_{\,\,\gamma} ,
\label{BR} \\
&& \pi^\rho_{\,\,\sigma}
\pi^\nu_{\,\,\lambda} \nabla_{[\rho} Q^\mu_{\,\,\nu]} =0  .
\eer{integ}
We also showed that they make up a complete set, i.e.\ there are
no further independent conditions to be found.

It may seem that these conditions are not in fact conditions on the
worldsheet fields, which we claimed we were after. However, they are
explicit conditions on the structures on the brane, which is
equivalent to restrictions of the motion of the string endpoints.

The first condition says that the boundary map $R^\mu_{\,\,\nu}$
preserves the spacetime metric. This implies that the metric
diagonalises so that it consists of a Neumann-Neumann part and a
Dirichlet-Dirichlet part, but has no ``mixed'' components.  One may
see how this happens by going to adapted coordinates; there we have
seen that $R^\mu_{\,\,\nu}$ has the diagonal form (\ref{Rmatrix}), so
it follows immediately from (\ref{RgR}) that the metric must be
block-diagonal, $G_{\mu\nu} = \diag(G_{mn}, G_{ij})$.

There are two things to note here. First, the diagonal form of the
metric does not necessarily imply that the two constituent blocks are
completely decoupled. There may still be interdependence such that
e.g.\ $G_{mn}$ depends on Dirichlet coordinates.  Second, it is
important to realise that the condition (\ref{RgR}) is not imposing
restrictions on the spacetime metric, but on the D-brane --- its
embedding in target space must be such that the metric diagonalises.

Moving on to the second boundary condition, Eq.\ (\ref{BR}), we see
that it is a condition on the Neumann-Neumann part of
$R^\mu_{\,\,\nu}$.  It is essentially the definition of the B-field
mentioned in Section~\ref{ansatz}; in adapted coordinates it
schematically looks like $E^{-1} E^T =R$ (along the Neumann-Neumann
directions). From this form we can derive the behaviour of the string
for large B-field; as we take $B \rightarrow \infty$, the
Neumann-Neumann block of $R^\mu_{\,\,\nu}$ goes to
$-\delta^m_{\,\,n}$, i.e.\ we find that the Neumann conditions turn
into Dirichlet conditions. That is, the endpoint of the string is
completely frozen in all directions in this limit.  It is also
interesting to note that, in the limit $B \rightarrow 0$, we find
$R^\mu_{\,\,\nu} \rightarrow r^\mu_{\,\,\nu}$.

Last but not least, we want to interpret the condition (\ref{integ}).
This is actually a precise mathematical statement saying that the
projector $\pi^\mu_{\,\,\nu}$ is \emph{integrable} \cite{Yano}. There
are several ways of understanding the content of this statement
\cite{YanoKon}, but perhaps the most intuitive one is to consider two
displacements along the Neumann directions. If the commutator
$[\delta_1, \delta_2] X^\mu$ between them vanishes, then regardless of
which order we do them in, we always end up in the same point. If, on
the other hand, they do not commute, the final position \emph{will}
depend on the order of displacement. The latter case corresponds to
some singular D-brane about which we cannot say much.  The former
situation, however, is the interesting one, because it occurs if and
only if (\ref{integ}) holds, as is easily seen by inserting $\delta
X^\mu = \pi^\mu_{\,\,\nu} \delta X^\nu$ in the commutator.  It implies
local smoothness at the point $u$ in the sense that we can extend the
basis of $\pi_u$ to a neighbourhood of $u$ \cite{Reinhart}. In
particular the adapted coordinate basis, where $\pi^\mu_{\,\,\nu}$
takes the form
\beq
\pi^\mu_{\,\,\nu}= \left(
\begin{array}{cc}
\delta^m_{\,\,n} &  0\\
0 & 0
\end{array}
\right) ,
\eeq{pimatrix}
can be extended to a neighbourhood of $u$ so that $\pi^\mu_{\,\,\nu}$
is given by (\ref{pimatrix}) on the whole neighbourhood.\footnote{Note
  that this extension is only possible along the Neumann directions.
  If $Q^\mu_{\,\,\nu}$ were integrable too, then we would be able to
  extend the basis in all directions.}  And since we can define
$Q^\mu_{\,\,\nu}$ wherever $\pi^\mu_{\,\,\nu}$ is defined, and hence
$r^\mu_{\,\,\nu}$, it follows that we may extend all three of our
(1,1)-tensors to a neighbourhood of $u$.

Another consequence of $\pi$-integrability is that the Neumann-Neumann
part of the metric is independent of the Dirichlet directions
\cite{YanoKon} (though as long as $Q^\mu_{\,\,\nu}$ is not integrable,
the Dirichlet-Dirichlet part of $G_{\mu\nu}$ may still depend on the
Neumann coordinates).

This is as far as we get in our interpretation of the conditions
(\ref{RgR})--(\ref{integ}) without making additional assumptions.  But
the properties of our structures $Q^\mu_{\,\,\nu}$,
$\pi^\mu_{\,\,\nu}$ and $r^\mu_{\,\,\nu}$ are exactly the same as
those of the structures associated with \emph{almost product
  manifolds} \cite{YanoKon}, except that they are not globally
defined. We therefore found it irresistible to add this extra property
in Papers~II and III, and investigate the consequences.

\section{Globally defined conditions}
\label{Global}

By ``globally defined'' we mean that our structures $Q^\mu_{\,\,\nu}$,
$\pi^\mu_{\,\,\nu}$ and $r^\mu_{\,\,\nu}$, as well as
$R^\mu_{\,\,\nu}$, all are defined not only at a point $u$ (and, by
integrability, on a neighbourhood of $u$), but at every point $v$
in spacetime.

In this case, $\pi^\mu_{\,\,\nu}$ and $Q^\mu_{\,\,\nu}$ are
\emph{distributions} on $\M$ \cite{Yano}. That is, they assign a space
of vectors ($\pi_v$ and $Q_v$, respectively) to each point $v \in \M$,
thus splitting the whole tangent space $T(\M)$ (not just the tangent
plane at a point $u$, cf.\ Eq.\ (\ref{Tsplit})) into a direct sum,
\beq
T (\M) = \pi \oplus Q .
\eeq{Tsplit2}
The tensor $r^\mu_{\,\,\nu}$ is now an \emph{almost product
  structure}, defining the target space as an \emph{almost product
  manifold} \cite{YanoKon}. The word ``almost'' refers to the fact
that the manifold is not necessarily a direct product, despite its
tangent space being a direct sum.\footnote{The failure of the split
  (\ref{Tsplit2}) to split $\M$ into a direct product is measured by
  the \emph{Nijenhuis tensor} $N^\rho_{\,\,\mu\nu} \equiv
  r^\gamma_{\,\,\mu} r^\rho_{\,\,[\nu,\gamma]} - r^\gamma_{\,\,\nu}
  r^\rho_{\,\,[\mu,\gamma]}$ \cite{Borowiec}.} An example of such a
manifold is a locally product manifold, i.e.\ a manifold that locally
looks like a direct product but globally is not (think of the M\"obius
strip).

Let us have a fresh look at the boundary conditions
(\ref{RgR})--(\ref{integ}) through our ``global glasses.''
We saw in Section~\ref{BCs} that the first condition implies that the
metric diagonalises. In adapted coordinates it is then easily seen
that the metric preserves also the almost product structure, since the
latter takes the form
$$
r^\mu_{\,\,\nu}= \left(
\begin{array}{cc}
\delta^m_{\,\,n} &  0\\
0 & -\delta^i_{\,\,j}
\end{array}
\right) 
$$
in this basis. Thus the spacetime manifold is an almost product
manifold with a (pseudo-) Riemannian metric that is preserved by the
almost product structure; such a manifold is called a (pseudo-)
Riemannian almost product manifold \cite{YanoKon}.

The second boundary condition, Eq.\ (\ref{BR}), gives no new
information, so we turn directly to the condition (\ref{integ}), which
says that $\pi^\mu_{\,\,\nu}$ is an integrable distribution. Such an
object is by definition a \emph{foliation} \cite{Reinhart,Tondeur}.
This means that, at every point $v$ in spacetime, $\pi^\mu_{\,\,\nu}$
locally defines, on a neighbourhood of $v$, a set of hyperplanes, or
\emph{leaves}, that in adapted coordinates are described by
the system of equations $\{X^i$=
constant$\}$ \cite{Tondeur}. These hyperplanes are contained in the
space $\pi_v$, but a priori need not be of the same dimension, so they
are not necessarily D-branes (recall that $\pi^\mu_{\,\,\nu}$ defines
the dimensionality of the D-brane).  However, there is a theorem by
Frobenius saying that when $\pi^\mu_{\,\,\nu}$ is a foliation,
then through every point $v \in \M$ there is a unique
submanifold of $\M$ whose tangent space is $\pi_v$ (called a
\emph{maximal integral manifold} of $\pi^\mu_{\,\,\nu}$)
\cite{Boothby}; this is the D-brane we were looking for.

In conclusion, globally defined superconformal boundary conditions
imply that spacetime is foliated by D-branes which are submanifolds
embedded in such a way that spacetime may be viewed as a (pseudo-)
Riemannian almost product manifold.  Let us give some concrete
examples of such D-brane embeddings.

\subsection{Examples}

As we mentioned above, the obvious special case of an almost product
manifold is a direct product, e.g.\ a torus (a direct product of
one-cycles). Thus, if spacetime were a torus, the superconformally
allowed D-branes would be those that wrap one or more of the
constituent one-cycles.

More interestingly, many exact solutions of Einstein's equations are
almost product manifolds, e.g.\ the Schwarzschild and Robertson-Walker
spaces. To be explicit, we take a closer look at the Schwarzschild
space; it makes an instructive example of the way different D-brane
embeddings determine the type of the spacetime manifold.

The Schwarzschild metric is
\beq
ds^2 = -f(r) dt^2 + h(r) dr^2 + r^2 d\Omega^2 ,
\eeq{Schwarzschild}
where $t$ is time, $r$ is the radial coordinate, $f(r) \equiv 1-M/r$
with $M$ a constant, $h(r) \equiv 1/f(r)$, and $d\Omega^2$ is the
metric on a two-sphere.  Assuming that $t$ is the time coordinate also
on the worldvolume of the D-brane, there is only one embedding which
is allowed by the superconformal boundary conditions
(\ref{RgR})--(\ref{integ}).  We can write the manifold at hand as a
\emph{warped product manifold}
$$
( \, \M_1 \times \M_2 \, , \, G_1(t,r) \oplus r^2 G_2(\Omega^2) \,) ,
$$
with $G_1(t,r) \equiv -f(r) dt^2 + h(r) dr^2$ being the metric on
$\M_1$ and $G_2(\Omega^2) \equiv d\Omega^2$ the metric on $\M_2$. It
is called ``warped'' because the factor in front of $G_2$ (in our case
$r^2$) depends on the coordinates of $\M_1$. Then the brane can extend
along $t$ and $r$, i.e.\ we have a D1-brane that coincides with the
manifold $\M_1$. Here $G_1(t,r)$ is the metric on the brane; it is
independent of the $\M_2$-coordinates, which is consistent with
integrability of $\pi^\mu_{\,\,\nu}$. On the other hand, there is a
problem with integrability at the singularity $r=0$; nevertheless, the
general principle should be clear.

Alternatively, we may write the metric (\ref{Schwarzschild}) as a
different almost product manifold,
$$
( \, \M_1 \times \M_2 \, , \, G_1(t,r,\Omega^2) \oplus G_2(r) \, ) ,
$$
with $G_1(t,r,\Omega^2) \equiv -f(r) dt^2 + r^2 d\Omega^2$ and $G_2(r)
\equiv h(r) dr^2$. Here we could have a D2-brane wrapping $\M_1$.
However, it would not be superconformal, since $\pi^\mu_{\,\,\nu}$ is not
integrable in this case; the Neumann-Neumann metric
$G_1(t,r,\Omega^2)$ depends on the Dirichlet direction $r$.
Finally, there is also the possibility of a D0-brane, extending only
along $t$, in which case spacetime again is a warped product manifold,
with warp factor $-f(r)$. Also this brane breaks superconformal
invariance, since the metric on the brane depends on the Dirichlet
directions.

\begin{figure}[ht]
  \epsfxsize=3cm
  \centerline{\epsfbox{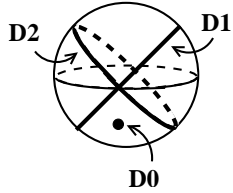}}
  \caption{\footnotesize Various D-brane embeddings
in a Schwarzschild space. The Schwarzschild space is represented here by
a two-sphere, and the D-branes
are, respectively, a D2-brane wrapping the two-sphere part of the
Schwarzschild space (represented here by a circle), a D1-brane
extending along the radius, and a D0-brane, which is just a point
in the spacelike directions. Only the D1-brane is allowed
by the globally defined
superconformal boundary conditions (\ref{RgR})--(\ref{integ}).}
  \label{Embeddings}
\end{figure}

The various embeddings are illustrated in Fig.~\ref{Embeddings}.

\section{The WZW model}
\label{WZW}

In the preceding sections we have learnt the minimal requirements for
a D-brane to preserve superconformal symmetry in a general spacetime.
To put our boundary conditions to the test, we want to study a special
case, i.e.\ a string propagating on a given manifold. The most obvious
choice is the extensively studied \emph{WZW model} \cite{Gawedzki}.
This is a nonlinear sigma model defined on a \emph{group manifold} of
some Lie group $G$; that is, the target space coordinates transform
into each other under a group $G$.  The importance of this
model is due to the fact that it is exactly solvable, and provides a
tractable setting for studying D-branes in curved backgrounds.  In
particular, when $G$ is a semisimple Lie group, the background fields
simplify in such a way as to make our boundary conditions more
transparent.

\subsection{Symmetries}

The target space (and the action) is invariant under
$G \times G$ transformations, so there are two chiral currents (i.e.\ 
each current depends only on one of the worldsheet coordinates
$\xi^{\pm}$) associated with this symmetry, one for each group $G$.
They can be derived from the superspace action (\ref{sigmasusy}) by
varying it with respect to the ``left'' and ``right'' group,
respectively. The result is left- and right-moving currents,
$$
J_{+} \equiv -\mathbf{l}_{\,\,\mu} D_+ \Phi^\mu, \quad\quad
J_{-} \equiv \mathbf{r}_{\,\,\mu} D_- \Phi^\mu ,
$$
where $\mathbf{l}_{\,\,\mu}$ and $\mathbf{r}_{\,\,\mu}$
are the superfield Killing vectors associated with each symmetry.
We denote the lowest (fermionic) components of $J_{\pm}$ by
\beq
j_{+} \equiv J_{+}| = -l_\mu \psi_+^\mu, \quad\quad
j_{-} \equiv J_{-}| = r_\mu \psi_-^\mu ,
\eeq{jpm}
where $l_\mu$ and $r_\mu$ are the lowest components of the superfield
Killing vectors.\footnote{The vector $r_\mu$ should not be confused
  with the almost product structure $r^\mu_{\,\,\,\nu}$ in
  Section~\ref{Global}.}

The vectors $l^\mu$ and $r^\mu$ can be expanded in the Lie algebra
basis $\{ T^A \}$ of their corresponding groups, as $l^\mu \equiv
l^\mu_{\,\,A} T^A$ and $r^\mu \equiv r^\mu_{\,\,A} T^A$. Then the
coefficients $l^\mu_{\,\,A}$ and $r^\mu_{\,\,A}$ satisfy
the corresponding Lie algebra,
$$
[l_A , l_B] = -f_{AB}^{\,\,\,\,\,\,\,\,\,C} l_C,
\quad\quad
[r_A , r_B] = f_{AB}^{\,\,\,\,\,\,\,\,\,C} r_C
$$
(the sign is just a convention),
and they commute with each other, $[l_A , r_B] = 0$.
Here the Lie bracket for vectors $v^\mu_{\,\,A}$ is defined as
$[v_A , v_B]^\mu \equiv  v^\nu_{\,\,A}
\d_{\nu} v^\mu_{\,\,B} - v^\nu_{\,\,B} \d_{\nu} v^\mu_{\,\,A}$.

\subsection{The gluing map}

From (\ref{jpm}) follows immediately that the fermionic ansatz
(\ref{fermbc}) translates into a boundary condition for the group
currents,
$$
j^A_- = \eta \, R^A_{\,\,\,B} \, j^B_+ ,
$$
where
\beq
R^A_{\,\,\,B} \equiv  -l^A_{\,\,\mu} R^\mu_{\,\,\,\nu} r^\nu_{\,\,B} .
\eeq{Rdef}
The transformation in (\ref{Rdef}) is essentially a change of
basis from spacetime vectors to the Lie algebra basis \cite{Stanciu}.

The object $R^A_{\,\,\,B}$ is a gluing map between the chiral currents
at the worldsheet boundary. Since it maps $j^B_+$ to $j^A_-$, which
are elements of the Lie algebra, it is clearly a map from the Lie
algebra into itself. What more we can say about this map?  It is
usually assumed in the literature to be a constant \emph{Lie algebra
  automorphism} \cite{Stanciu,Alekseev,Itoh}. This means that it
preserves the Lie algebra structure in the sense that
\beq
\left[ R^C_{\,\,\,A} T_C, R^D_{\,\,\,B} T_D \right]
=  f_{AB}^{\,\,\,\,\,\,\,\,\,C} R^F_{\,\,\,C} T_F .
\eeq{Lieaut}
However, this restriction is always made for practical reasons, as a
nonconstant gluing map makes the quantum theory unwieldy. And indeed
we found in Paper~IV that $R^A_{\,\,\,B}$ in the $\N$=1 model can in
principle be a much more general object.

To see how we reach this conclusion, let us write down the boundary
conditions for the WZW model. For this purpose, we take advantage of
the fact that on group manifolds the metric and torsion can be
expressed entirely in terms of Killing vectors, Lie algebra structure
constants and the Cartan-Killing metric\footnote{The Cartan-Killing
  metric is defined as $\eta_{AB} \equiv -
  f_{AC}^{\,\,\,\,\,\,\,\,\,D} f_{BD}^{\,\,\,\,\,\,\,\,\,C}$.}
$\eta_{AB}$ \cite{HullSpence}.
As a consequence, the boundary conditions (\ref{RgR}) and
(\ref{integ}) can be written as equations involving only these
quantities plus $R^A_{\,\,\,B}$, and we end up with the following
conditions,
\beq
\eta_{AB} = R^C_{\,\,\,A} \eta_{CD} R^D_{\,\,\,B} ,
\eeq{RetaR}
\beq
f_{ABC} - f_{DEF}
R^D_{\,\,\,A} R^E_{\,\,\,B} R^F_{\,\,\,C} =
 \eta_{DE} R^D_{\,\,\,[A} \left( \L_C R^E_{\,\,\,B]} \right) .
\eeq{Gcond2}
Here the Lie derivative is defined as
$$
\L_{C} R^E_{\,\,\,B} \equiv \L_{r_C} R^E_{\,\,\,B} -
R^H_{\,\,\,C} \L_{l_H} R^E_{\,\,\,B}
= \left( r^\mu_C - R^H_{\,\,\,C} l^\mu_H \right) R^E_{\,\,\,B,\mu},
$$
with $\L_{r_C}$ the usual Lie derivative with respect
to the vector $r^\mu_{\,\,C}$.

The first condition, Eq.\ (\ref{RetaR}), says that the gluing map
preserves the Cartan-Killing metric. But it is the second condition,
Eq.\ (\ref{Gcond2}), that interests us. To see what it means,
let us first use (\ref{RetaR}) to manipulate the Lie bracket
(\ref{Lieaut})
a little. We can rewrite the left-hand side of (\ref{Lieaut}) as
$$
R^D_{\,\,\,A}  R^E_{\,\,\,B} \left[ T_D, T_E \right]
 = R^D_{\,\,\,A}  R^E_{\,\,\,B} f_{DE}^{\,\,\,\,\,\,\,\,\,F} T_F ,
$$
whence, using (\ref{RetaR}),
follows that $R^A_{\,\,\,B}$ is a Lie algebra automorphism if and
only if 
$$
f_{ABC} -  f_{DEF} R^D_{\,\,\,A} R^E_{\,\,\,B} R^F_{\,\,\,C} =0.
$$
Thus we see that $R^A_{\,\,\,B}$ is a Lie algebra automorphism if and
only if the right-hand side of (\ref{Gcond2}) vanishes.

Hence we draw two conclusions from (\ref{Gcond2}). First,
$R^A_{\,\,\,B}$ does not have to be a Lie algebra automorphism.
Second, it can be a Lie algebra automorphism without being constant,
since the right-hand side can vanish even if $R^A_{\,\,\,B}$ is
nonconstant.  The interpretation of this condition in terms of
D-branes was not addressed in Paper~IV, but it is known that for the
special case where $R^A_{\,\,\,B}$ is constant, conformally invariant
D-branes are obtained as conjugacy classes of the group $G$
\cite{Alekseev}.

\backmatter


\addcontentsline{toc}{chapter}{Bibliography}

\begin{thebibliography}{6666}
%
\bibitem{Bilal2}
A.~Bilal,
\emph{Higher-derivative corrections to the non-abelian Born-Infeld action},
Nucl.\ Phys.\ {\bf B618} (2001) 21-49,
hep-th/0106062.
%%CITATION = HEP-TH 0106062;%%
%
\bibitem{Koerber}
P.~Koerber and A.~Sevrin,
\emph{The non-abelian D-brane effective action through order $\alpha'{}^4$},
JHEP {\bf 0210} (2002) 046,
hep-th/0208044.
%%CITATION = HEP-TH 0208044;%%
%
\bibitem{Greene}
B.~R.~Greene,
\emph{String Theory on Calabi-Yau Manifolds},
TASI lecture notes (1996),
hep-th/9702155.
%%CITATION = HEP-TH 9702155;%%
%
\bibitem{Aspinwall1}
P.~S.~Aspinwall,
\emph{K3 surfaces and string duality},
TASI lecture notes (1996),
hep-th/9611137.
%%CITATION = HEP-TH 9611137;%%
%
\bibitem{PolII}
J.~Polchinski, \emph{String theory}, Vol II
(Cambridge University Press, 1998).
%
\bibitem{IS}
K.~Intriligator and N.~Seiberg,
\emph{Mirror Symmetry in Three Dimensional Gauge Theories},
Phys.\ Lett.\ {\bf B387} (1996) 513-519,
hep-th/9607207.
%%CITATION = HEP-TH 9607207;%%
%
\bibitem{Kronheimer}
P.~B.~Kronheimer,
\emph{The construction of ALE spaces as Hyper-K\"ahler
quotients},
J.~Diff.\ Geom.\ {\bf 29} (1989) 665.
%
\bibitem{E6}
U.~Lindstr\"om, M.~Ro\v{c}ek and R.~von~Unge,
\emph{Hyperk\"ahler quotients and algebraic curves}
JHEP {\bf 0001} (2000) 022,
hep-th/9908082.
%%CITATION = HEP-TH 9908082;%%
%
\bibitem{E7}
I.~Y.~Park and R.~von~Unge,
\emph{Hyperk\"ahler quotients, mirror symmetry, and F-theory},
JHEP {\bf 0003} (2000) 037,
hep-th/0001051.
%%CITATION = HEP-TH 0001051;%%
%
\bibitem{Fulton}
W.~Fulton and J.~Harris, \emph{Representation Theory: A First Course}
(Springer-Verlag, 1991).
%
\bibitem{Ryder}
L.~H.~Ryder, \emph{Quantum field theory}
(Cambridge University Press, 1996).
%
\bibitem{Maggiore}
M.~Maggiore, \emph{Champs et Particules}, Chapter 9,
lecture notes 2001/02, D\'epartement de Physique Th\'eorique
de l'Universit\'e de Gen\'eve.
%
\bibitem{Bilal}
A.~Bilal,
\emph{Duality in $\N$=2 susy $SU(2)$ Yang-Mills theory: A pedagogical introduction to the work of Seiberg and Witten},
hep-th/9601007.
%%CITATION = HEP-TH 9601007;%%
%
\bibitem{Johnson}
C.~V.~Johnson and R.~C.~Myers,
\emph{Aspects of Type IIB theory on ALE spaces},
Phys.\ Rev.\ {\bf D55} (1997) 6382-6393,
hep-th/9610140.
%%CITATION = HEP-TH 9610140;%%
%
\bibitem{Wess}
J.~Wess and J.~Bagger, \emph{Supersymmetry and supergravity}
(Princeton University Press, 1992).
%
\bibitem{Aspinwall2}
P.~S.~Aspinwall,
\emph{Compactification, geometry and duality: N=2},
TASI lecture notes (1999),
hep-th/0001001.
%%CITATION = HEP-TH 0001001;%%
%
\bibitem{HKLR}
N.~J.~Hitchin, A.~Karlhede, U.~Lindstr\"om and M.~Ro\v{c}ek,
\emph{Hyperk\"ahler metrics and supersymmetry},
Commun.\ Math.\ Phys.\ {\bf 108} (1987) 535-589.
%
\bibitem{Lerche}
W.~Lerche,
\emph{Introduction to Seiberg-Witten theory and its stringy origin},
Nucl.\ Phys.\ Proc.\ Suppl.\ {\bf 55B} (1997) 83-117;
Fortsch.\ Phys.\ {\bf 45} (1997) 293-340,
hep-th/9611190.
%%CITATION = HEP-TH 9611190;%%
%
\bibitem{SW1}
N.~Seiberg and E.~Witten,
\emph{Monopole condensation and confinement in N=2 supersymmetric
Yang-Mills theory},
Nucl.\ Phys.\ {\bf B426} (1994) 19-52;
Erratum-ibid.\ {\bf B430} (1994) 485-486,
hep-th/9407087.
%%CITATION = HEP-TH 9407087;%%
%
\bibitem{Argyres1}
P.~C.~Argyres, M.~R.~Plesser and N.~Seiberg,
\emph{The moduli space of N=2 SUSY QCD and duality in N=1 SUSY QCD},
Nucl.\ Phys.\ {\bf B471} (1996) 159-194,
hep-th/9603042.
%%CITATION = HEP-TH 9603042;%%
%
\bibitem{Koblitz}
N.~Koblitz, \emph{Introduction to elliptic curves and
modular forms}, 2nd ed.\ (Springer-Verlag, 1993).
%
\bibitem{Kodaira}
K.~Kodaira,
\emph{On compact analytic surfaces, II-III},
Ann.\ of Math.\ {\bf 77} (1963) 563; {\bf 78} (1963) 1.
%
\bibitem{Argyres2}
P.~C.~Argyres, M.~R.~Plesser, N.~Seiberg and E.~Witten,
\emph{New N=2 superconformal field theories in four dimensions},
Nucl.\ Phys.\ {\bf B461} (1996) 71,
hep-th/9511154.
%%CITATION = HEP-TH 9511154;%%
%
\bibitem{Eguchi}
T.~Eguchi, K.~Hori, K.~Ito and S.-K.~Yang,
\emph{Study of N=2 superconformal field theories in 4 dimensions},
Nucl.\ Phys.\ {\bf B471} (1996) 430,
hep-th/9603002.
%%CITATION = HEP-TH 9603002;%%
%
\bibitem{Minahan1}
J.~Minahan and D.~Nemeschansky,
\emph{An N=2 superconformal fixed points with $E_6$ global symmetry},
Nucl.\ Phys.\ {\bf B482} (1996) 142-152,
hep-th/9608047.
%%CITATION = HEP-TH 9608047;%%
%
\bibitem{Minahan2}
J.~Minahan and D.~Nemeschansky,
\emph{Superconformal fixed points with $E_n$ global symmetry},
Nucl.\ Phys.\ {\bf B489} (1997) 24-46,
hep-th/9610076.
%%CITATION = HEP-TH 9610076;%%
%
\bibitem{Noguchi}
M.~Noguchi, S.~Terashima and S.-K.~Yang,
\emph{N=2 superconformal field theory with ADE global symmetry
on a D3-brane probe},
Nucl.\ Phys.\ {\bf B556} (1999) 115-151,
hep-th/9903215.
%%CITATION = HEP-TH 9903215;%%
%
\bibitem{Seiberg}
N.~Seiberg,
\emph{Five dimensional SUSY field theories, non-trivial
fixed points and string dynamics},
Phys.\ Lett.\ {\bf B388} (1996) 753-760,
hep-th/9608111.
%%CITATION = HEP-TH 9608111;%%
%
\bibitem{Lerche2}
W.~Lerche and N.~P.~Warner,
\emph{Exceptional SW geometry from ALE fibrations},
Phys.\ Lett.\ {\bf B423} (1998) 79-86,
hep-th/9608183.
%%CITATION = HEP-TH 9608183;%%
%
\bibitem{Vafa}
C.~Vafa,
\emph{Evidence for F-Theory},
Nucl.\ Phys.\ {\bf B469} (1996) 403-418,
hep-th/9602022.
%%CITATION = HEP-TH 9602022;%%
%
\bibitem{Johansen}
A.~Johansen,
\emph{A comment on BPS states in F-theory in 8 dimensions},
Phys.\ Lett.\ {\bf B395} (1997) 36,
hep-th/9608186.
%%CITATION = HEP-TH 9608186;%%
%
\bibitem{Gaberdiel}
M.~R.~Gaberdiel and B.~Zwiebach,
\emph{Exceptional groups from open strings},
Nucl.\ Phys.\ {\bf B518} (1998) 151-172,
hep-th/9709013.
%%CITATION = HEP-TH 9709013;%%
%
\bibitem{Banks}
T.~Banks, M.~Douglas and N.~Seiberg,
\emph{Probing F-theory with branes},
Phys.\ Lett.\ {\bf B387} (1996) 278,
hep-th/9605199.
%%CITATION = HEP-TH 9605199;%%
%
\bibitem{Ansar2}
A.~Fayyazuddin and M.~Spalinski,
\emph{The Seiberg-Witten differential from M-Theory},
Nucl.\ Phys.\ {\bf B508} (1997) 219-228,
hep-th/9706087.
%%CITATION = HEP-TH 9706087;%%
%
\bibitem{Aharony}
O.~Aharony, A.~Fayyazuddin and J.~Maldacena,
\emph{The large N limit of ${\cal N} =2,1$ field theories from threebranes in F-theory},
JHEP {\bf 9807} (1998) 013,
hep-th/9806159.
%%CITATION = HEP-TH 9806159;%%
%
\bibitem{Sen}
A.~Sen,
\emph{F-theory and orientifolds},
Nucl.\ Phys.\ {\bf B475} (1996) 562-578,
hep-th/9605150.
%%CITATION = HEP-TH 9605150;%%
%
\bibitem{Feng}
B.~Feng, A.~Hanany, Y.-H.~He and N.~Prezas,
\emph{Stepwise projection: toward brane setups for generic
orbifold singularities},
JHEP {\bf 0201} (2002) 040,
hep-th/0012078.
%%CITATION = HEP-TH 0012078;%%
%
\bibitem{Douglas}
M.~Douglas and G.~Moore,
\emph{D-branes, quivers, and ALE instantons},
hep-th/9603167.
%%CITATION = HEP-TH 9603167;%%
%
\bibitem{PolI}
J.~Polchinski, \emph{String theory}, Vol I
(Cambridge University Press, 1998).
%
\bibitem{Mumford}
D.~Mumford and J.~Fogarty, \emph{Geometric invariant theory}
(Springer, 1982).
%
\bibitem{ADEproc}
C.~Albertsson, B.~Brinne, U.~Lindstr\"om, M.~Ro\v{c}ek and R.~von~Unge,
\emph{ADE-quiver theories and mirror symmetry},
Nucl.\ Phys.~B Proc.\ Suppl.\ {\bf 102} \& {\bf 103} (2001) 3-10,
hep-th/0103084.
%%CITATION = HEP-TH 0103084;%%
%
\bibitem{LZ1}
U.~Lindstr\"om and M.~Zabzine,
\emph{N=2 boundary conditions for non-linear sigma models and Landau-Ginzburg models},
JHEP {\bf 0302} (2003) 006,
hep-th/0209098.
%%CITATION = HEP-TH 0209098;%%
%
\bibitem{LZ2}
U.~Lindstr\"om and M.~Zabzine,
\emph{D-branes in N=2 WZW models},
to appear in Phys.\ Lett.\ B,
hep-th/0212042.
%%CITATION = HEP-TH 0212042;%%
%
\bibitem{Gawedzki}
K.~Gawedzki, \emph{Conformal field theory: a case study},
hep-th/9904145.
%%CITATION = HEP-TH 9904145;%%
%
\bibitem{Stanciu}
S.~Stanciu,
\emph{D-branes in group manifolds},
JHEP {\bf 0001} (2000) 025,
hep-th/9909163.
%%CITATION = HEP-TH 9909163;%%
%
\bibitem{GSWI}
M.~B.~Green, J.~H.~Schwarz and E.~Witten,
\emph{Superstring theory}, Vol I
(Cambridge University Press, 1987).
%
\bibitem{GSWII}
M.~B.~Green, J.~H.~Schwarz and E.~Witten, \emph{Superstring theory}, Vol II
(Cambridge University Press, 1987).
%
\bibitem{Bastianelli}
F.~Bastianelli and U.~Lindstr\"om,
\emph{Induced chiral supergravities in 2D},
Nucl.\ Phys.\ {\bf B416} (1994) 227,
hep-th/9303109.
%%CITATION = HEP-TH 9303109;%%
%
\bibitem{Ooguri}
H.~Ooguri, Y.~Oz and Z.~Yin,
\emph{D-branes on Calabi-Yau spaces and their mirrors},
Nucl.\ Phys.\ {\bf B477} (1996) 407-430,
hep-th/9606112.
%%CITATION = HEP-TH 9606112;%%
%
\bibitem{Alvarez}
E.~Alvarez, J.~L.~F.~Barbon and J.~Borlaf,
\emph{T-duality for open strings},
Nucl.\ Phys.\ {\bf B479} (1996) 218-242,
hep-th/9603089.
%%CITATION = HEP-TH 9603089;%%
%
\bibitem{Yano}
K.~Yano, \emph{Differential geometry on complex and almost complex spaces}
(Pergamon, Oxford, 1965).
%
\bibitem{YanoKon}
K.~Yano and M.~Kon, \emph{Structures on manifolds},
Series in Pure Mathematics, Vol 3
(World Scientific, Singapore, 1984).
%
\bibitem{Reinhart}
B.~L.~Reinhart, \emph{Differential geometry of foliations}
(Springer-Verlag, 1983).
%
\bibitem{Borowiec}
A.~Borowiec, M.~Ferraris, M.~Francaviglia and I.~Volovich,
\emph{Almost complex and almost product Einstein manifolds from a variational principle},
J.~Math.\ Phys.\ {\bf 40} (1999) 3446-3464,
dg-ga/9612009.
%%CITATION = DG-GA 9612009;%%
%
\bibitem{Tondeur}
P.~Tondeur, \emph{Foliations on Riemannian manifolds}
(Springer-Verlag, New York, 1988).
%
\bibitem{Boothby}
W.~M.~Boothby, \emph{An introduction to differentiable manifolds and Riemannian geometry}
(Academic Press, London, 1986).
%
\bibitem{Alekseev}
A.~Y.~Alekseev and V.~Schomerus,
\emph{D-branes in the WZW model},
Phys.\ Rev.\ {\bf D60} (1999) 061901,
hep-th/9812193.
%%CITATION = HEP-TH 9812193;%%
%
\bibitem{Itoh}
T.~Itoh and S.-J.~Sin,
\emph{A note on singular D-branes in group manifolds},
Nucl.\ Phys.\ {\bf B650} (2003) 497-521,
hep-th/0206238.
%%CITATION = HEP-TH 0206238;%%
%
\bibitem{HullSpence}
C.~M.~Hull and B.~Spence,
\emph{The gauged nonlinear sigma model with Wess-Zumino term},
Phys.\ Lett.\ {\bf B232} (1989) 204.
%
\end{thebibliography}
\end{document}